%% file: ENGLISH.TEX
\documentclass[
aps,%
12pt,%
final,%
notitlepage,%
oneside,%
onecolumn,%
nobibnotes,%
nofootinbib,%
superscriptaddress,%
noshowpacs,%
centertags]%
{revtex4}

\usepackage[english]{babel}
\usepackage{graphicx}

\newcommand\rem[1]{}

\input aadef.tex

\begin{document}

\begin{center}
{\Large\noindent
Properties of Contracting Massive Protostellar Cores
}

{\it L.\,E. Pirogov\,$^{*}$, P. M. Zemlyanukha, E. M. Dombek}

\scriptsize{Gaponov-Grekhov Institute of Applied Physics, Russian Academy of Sciences, Nizhny Novgorod, Russia} \\
\scriptsize{$^*$ e-mail: pirogov@appl.sci-nnov.ru}
\end{center}

\vskip 3mm

\noindent{\bf Abstract.} Studies of the structure and kinematics of cores 
associated with the regions of massive star 
and star cluster formation are necessary for constructing scenario 
for the evolution of these objects. 
We analyzed spectral maps of the massive cores of G012.418+00.506, 
G326.472+00.888, G328.567--00.535, G335.586--00.289 and G343.127--00.063 
from the MALT90 survey in the HCO$^+$(1--0) and H$^{13}$CO$^+$(1--0) lines. 
The cores are at different stages of evolution and have signs of contraction. 
By fitting spectral maps calculated within the framework of a spherically 
symmetric model into the observed ones, the parameters of the radial
profiles of density, turbulent velocity and contraction velocity were calculated. 
The power-law index of the density decay with distance from the center varies 
in the range of $\sim 1.5-2.8$. 
The lowest value is obtained for the core of G326.472+00.888 without 
internal sources. 
The contraction velocity in all cores depends weakly
on the distance from the center, decreasing with an index of $\sim 0.1$, 
which differs from the free-fall mode. 
There are indications of rotation for the cores of G328.567--00.535 
and G335.586--00.289. 
Analysis of $^{13}$CO(2--1) data from the SEDIGISM survey for the regions 
G012.418+00.506, G335.586--00.289, and G343.127--00.063 revealed motions 
from the surrounding gas toward the cores. 
The results obtained indicate that the massive cores under consideration 
interact with their environment and are apparently 
in a state of global collapse.

\vskip 2mm


\section{Introduction}

The scenario for the formation of massive stars
($\ga 8~M_{\odot}$) is under development [1, 2]. 
Of great importance is the question of the dynamic properties 
of dense cores where massive stars are formed and the
nature of the collapse. Theoretical works considering
the collapse of a spherically symmetric core predict
various solutions depending on the initial conditions [3]. 
Thus, a core initially in a state close to equilibrium,
with a small influence of external pressure, will
slowly evolve in a quasi-stationary regime to a nonequilibrium state. 
After a protostar is formed in the center of the core, the gas motions 
in its vicinity will be determined by the free-fall regime 
with a radial velocity profile $r^{-0.5}$. 
Over time, the region covered by the collapse increases. 
The isothermal singular sphere model [4] and the turbulent polytropic 
sphere model [5], proposed for cores in which massive stars are
formed, are examples of quasi-stationary solutions.
Nonequilibrium cores, unlike quasi-stationary ones,
being in a state of global collapse with a constant
velocity of motions directed from the outside inwards [3], 
will evolve significantly faster. 
The process of contraction
of a nonequilibrium core leads to fragmentation
and further growth of central objects, for example,
due to the mechanism of competitive accretion [6].
The model of global hierarchical collapse is based on this idea [7]. 
While the structure and nature of gas
motions in the inner region of the core, where free fall
occurs, are the same for both models, and the density
profiles in the outer regions have the form $r^{-2}$, 
it is the systematic velocity profile in the outer regions that can
be used to choose between the models.

Estimates of the systematic velocity of gas in the
core can be made from an analysis of the observed
molecular lines. 
Thus, an optically thick spectral line
formed in a contracting gas becomes asymmetric, and
its peak shifts to the ``blue'' side, which is associated
with differences in excitation conditions on the line of
sight and the Doppler effect. 
There is an analytical
method for estimating the contraction velocity on the
line of sight from the parameters of the observed line
possessing ``blue'' asymmetry (see, for example, [8]).
However, in order to estimate the radial dependence of
the contraction velocity in the core, calculations of
radiation transfer in the line within a certain model of
the core and fitting the calculated spectral map into
the observed one are required.

Fitting model spectral maps into observed ones
with simultaneous variation of the set of parameters
describing the structure and kinematics of the core is a
very time-consuming process even for fairly simple
spherically symmetric models due to the large number
of free parameters, correlations between them and
dependence on the initial conditions. 
To solve this
problem, an algorithm based on the use of a pre-calculated
array of spectral maps for a fairly wide range of
model parameters was developed in [9]. 
The algorithm
uses the principal component (PC) method to reduce
the dimensionality of the model and optimally fill the
parameter region containing the global minimum of
the error function. The k-nearest neighbor (kNN)
method is used to find the optimal parameter values
corresponding to the minimum.

Among the molecular lines, tracers of dense gas
($\ge 10^5$~cm$^{-3}$), the optically thick lines of the HCO$^+$ 
and HCN molecules are the most sensitive to the kinematics
and spatial distribution of the density (see, e.g., [10--12]). 
Using the spherically symmetric model and
the above algorithm, we analyzed the spectral maps of
the (1--0) transition of the HCO$^+$ and HCN molecules
and their rarer isotopes observed in the core of L1287,
in which a star cluster is forming [9], as well as the
HCO$^+$(1--0) and H$^{13}$CO$^+$(1--0) maps in the core of
G268.42--0.85, associated with the region of formation
of a high-mass star [13]. 
The values of the parameters
of the radial profiles of the density and velocity
were calculated. For both cores, the calculated radial
profiles of the contraction velocity turned out to be
different from those expected in the case of free fall,
indicating the preference of the model of an initially
nonequilibrium core for these objects.

In order to draw general conclusions about the
nature of gas contraction in the cores associated with
the regions of formation of massive stars and star clusters,
it is necessary to analyze a larger number of objects. 
For this purpose, objects with signs of contraction
were selected from the MALT90 catalog [14].
In this paper, we present estimates of the parameters of
the radial profiles of physical parameters, including
the systematic velocity, for five MALT90 objects using
an algorithm for fitting model spectral maps to
observed ones [9]. Section~2 provides brief information
about the MALT90 database and the sample of
objects. Section~3 describes the objects analyzed in
this paper. The results of the analysis of the maps in
molecular lines are given in Section~4, their discussion
is given in Section~5. The conclusions of the work are
formulated in Section~6.

\section{Selection of objects from the MALT90 database}
\label{sec:sources}

Surveys of regions of the galactic plane in the continuum
at submillimeter waves and in the far infrared
range (e.g. ATLASGAL [15], HiGal [16]), as well as in
the lines of the CO molecule and its isotopes (e.g.,
SEDIGISM [17], FUGIN [18]) provide important
information on the large-scale distribution of dust and gas. 
These surveys made it possible to identify the
cores of molecular dust clouds associated with the
regions of formation of high-mass stars and star clusters.
The MALT90 (The Millimetre Astronomy Legacy
Team 90~GHz) survey [14, 19], conducted in the
direction of massive dust clumps detected by the emission
in the continuum at a wavelength of 870~$\mu$m
(ATLASGAL [15]), represents the largest database of
spectroscopic observations of objects of this class.
\footnote{MALT90 data are publicly available in the Australian Telescope
Online Archive (ATOA), http://atoa.atnf.csiro.au}
The survey includes observations of more than
2000 objects conducted with the MOPRA-22m radio
telescope in the 3-mm wavelength range with an angular
resolution of $\sim 36''$. 
The observations were carried out in 16 spectral subranges. 
Most of the MALT90
objects are dense cores associated with different stages
of the formation of massive stars and star clusters.

As a result of a preliminary analysis of the MALT90
database, we identified 755 objects with noticeable
emission in the HCO$^+$(1--0) line (peak value $T_a^* \ge 1$~K). 
In 110 objects, the HCO$^+$(1--0) profiles have
noticeable asymmetry (the peak is shifted relative to
the center of the H$^{13}$CO$^+$(1--0) and N$_2$H$^+$(1--0) lines
to the ``blue'' or ``red'' side) or have a dip, while the
intensities of the ``blue'' and ``red'' peaks are close. In
20 cores with a dip, the amplitude of the ``blue'' peak
exceeds the amplitude of the ``red'' one. 
The presence of asymmetry of this type for optically thick 
HCO$^+$(1--0) lines and symmetric, close to Gaussian,
H$^{13}$CO$^+$(1--0) and N$_2$H$^+$(1--0) lines, the optical
depth of which is significantly lower, and the peak is
close to the position of the dip of the HCO$^+$(1--0)
lines, indicates probable contraction of the core [20].
These objects were selected for further evaluation of
physical parameters and analysis of the nature of gas
contraction.

This paper presents the results of the analysis of
observational data for five of the 20 selected cores at
different stages of evolution. The ``blue'' asymmetry of
the HCO$^+$(1--0) lines is observed at map scales larger
than the radiation pattern ($\ge 1'$). 
MALT90 data processing
consisted of extracting a section with a line
from the corresponding spectral subrange, subtracting
the 1st--2nd order baseline, and recalculating the
intensity scale from the antenna temperature to temperatures
reduced to the main beam of the radiation pattern. 
Since the G328.567--00.535 and G328.575--00.527 regions overlapped, 
they were averaged using the CLASS program from the GILDAS
\footnote{http://iram.fr/IRAMFR/GILDAS} 
package to increase the signal-to-noise ratio. 
The list of objects is given in Table 1. 
The names of the objects correspond
to the names of the regions from the MALT90 database.
Table 1 gives the galactic and equatorial coordinates
of the centers of the regions, distances to the
objects, their evolutionary status and associations with
other objects. Maps of the integrated intensities of
HCO$^+$(1--0), maps in the continuum according to
ATLASGAL data and the infrared maps according to
Spitzer data are shown in Fig.~1. The maps also show
maser sources and IRAS sources. A detailed description
of the objects is given in Section~3.

\begin{table}[p]
\setcaptionmargin{0mm}
\onelinecaptionsfalse
\captionstyle{flushleft}
\centering
\caption{List from cores from the MALT90 database}
\vskip 2mm
%\footnotesize
%\small
\scriptsize
\begin{tabular}{l|c|c|c|c|r|c|l}
\noalign{\hrule}\noalign{\smallskip}
Object           &$l$     &$b$    &$\alpha$(2000)  &$\delta$(2000) & $D$ & Evolutionary & Associations \\
                 &(deg)   &(deg)  &(hh:mm:ss)      &($^{o}:':''$)  & (kpc) & status     & with other objects \\
\noalign{\hrule}\noalign{\smallskip}
G012.418+00.506   &12.419   &0.507   & 18:10:51.1   &--17:55:49.6  & 1.8   & H   & IRAS 18079--1756, SIRDC \\
G326.472+00.888   &326.472  &0.889   & 15:42:29.6   &--53:58:26.7  & 2.5   & U   &  \\
G328.567--00.535  &328.568  &--0.534 & 15:59:37.4   &--53:45:51.7  & 2.9   & H   & IRAS 15557--5337, RCW99 \\
G335.586--00.289  &335.586  &--0.289 & 16:30:58.7   &--48:43:48.0  & 3.2   & A   & IRAS 16272--4837, SIRDC\_335.579-0.292 \\
G343.127--00.063  &343.128  &--0.063 & 16:58:17.5   &--42:52:04.0  & 2.8   & A   & IRAS 16547--4247, SIRDC , EGO\\
\noalign{\hrule}\noalign{\smallskip}
\end{tabular}
\flushleft
{\small
Kinematic distances to objects are taken from the survey [21]. 
The evolutionary status of cores according to the data of [19, 22] is as follows:
H -- H~II region, A -- protostellar core, U -- status not determined.
}
\label{table:list}
\end{table}

\begin{figure}[t!]
\setcaptionmargin{5mm}
\onelinecaptionsfalse
\captionstyle{flushleft}

\begin{minipage}[b]{0.38\textwidth}
    \includegraphics[width=\textwidth,angle=-0]{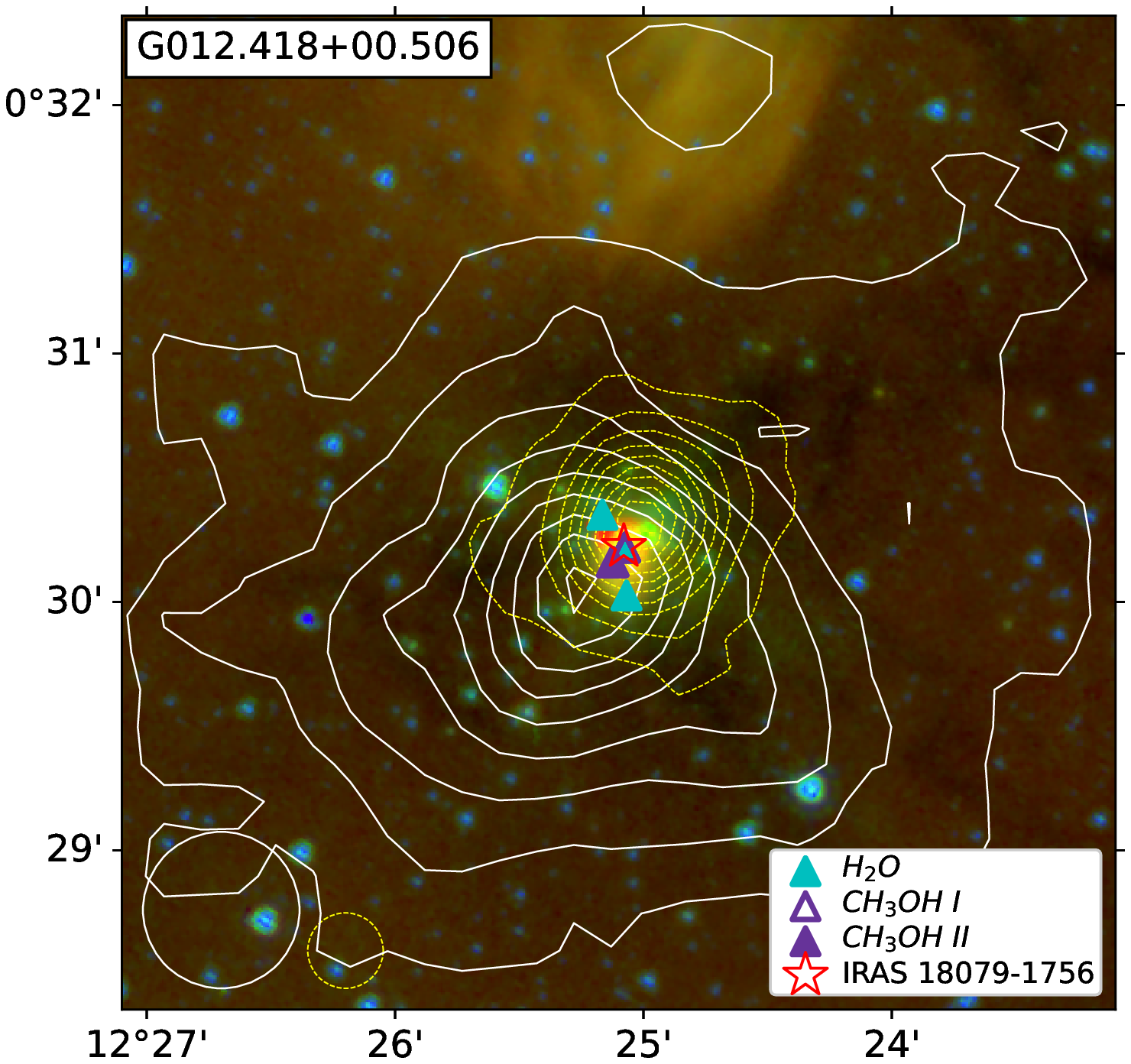}
\end{minipage}
\hspace{3mm}
\begin{minipage}[b]{0.38\textwidth}
    \includegraphics[width=\textwidth,angle=-0]{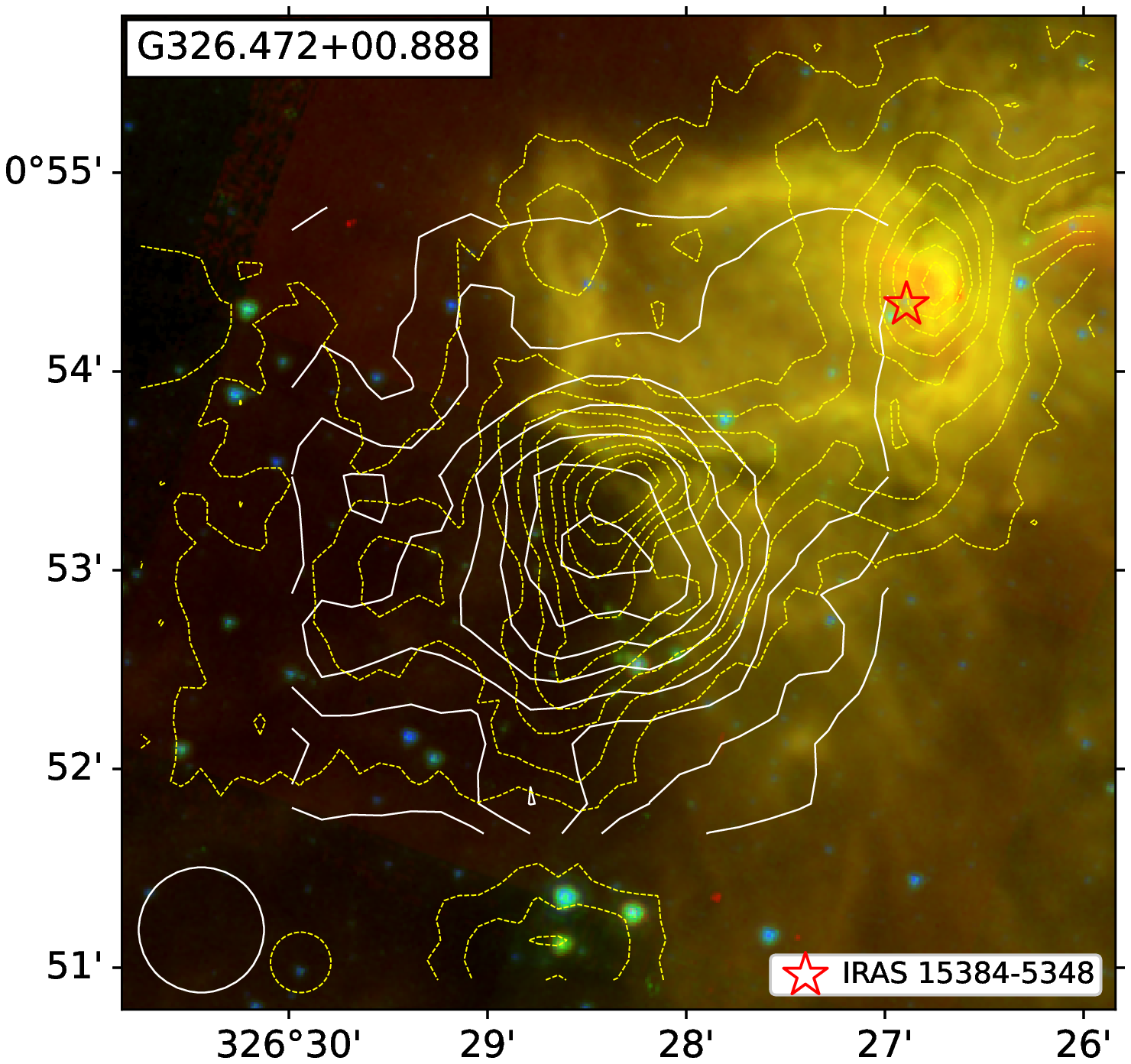}
\end{minipage}

\vspace{1mm}

\begin{minipage}[b]{0.38\textwidth}
    \includegraphics[width=\textwidth,angle=-0]{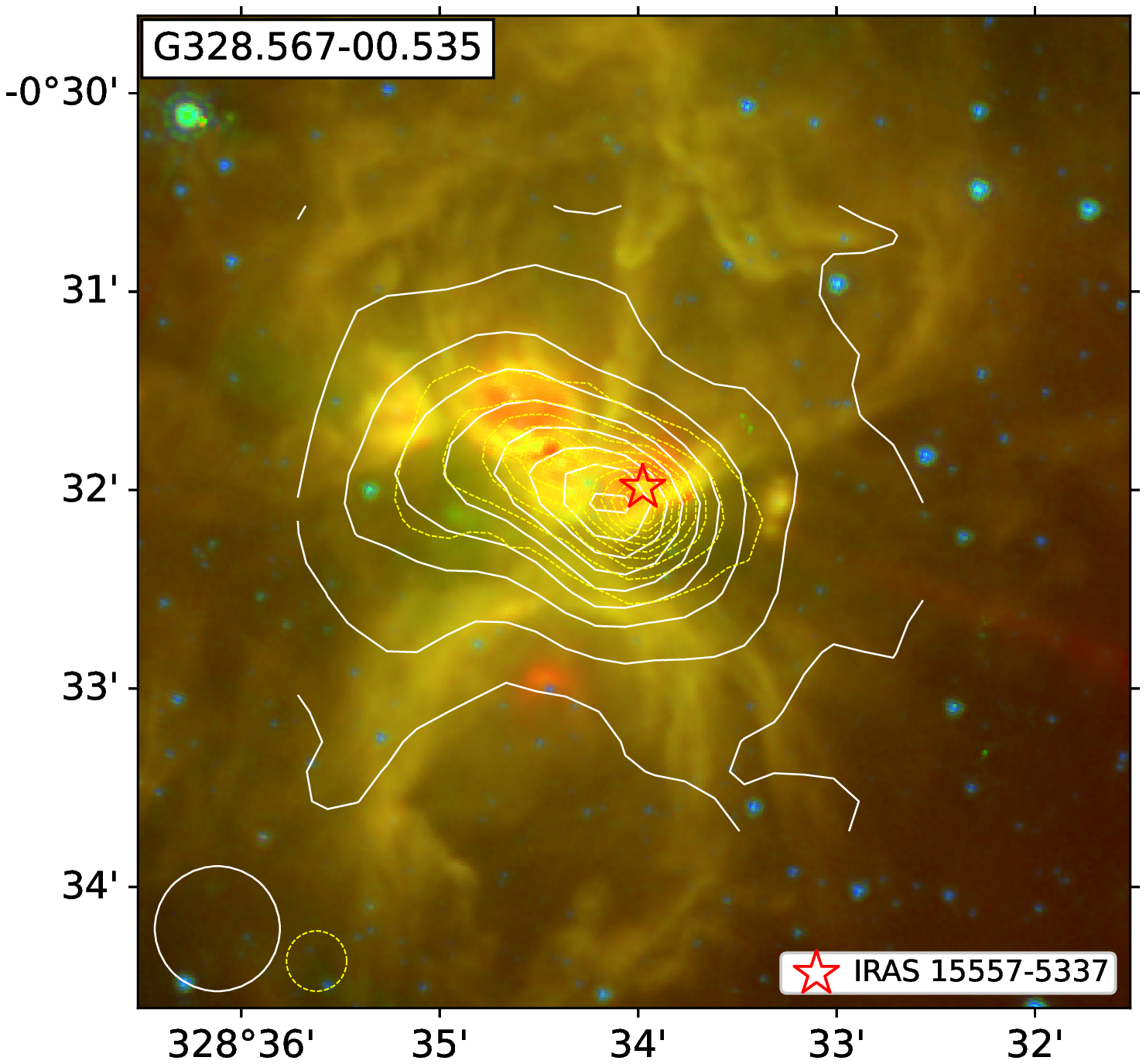}
\end{minipage}
\hspace{3mm}
\begin{minipage}[b]{0.38\textwidth}
    \includegraphics[width=\textwidth,angle=-0]{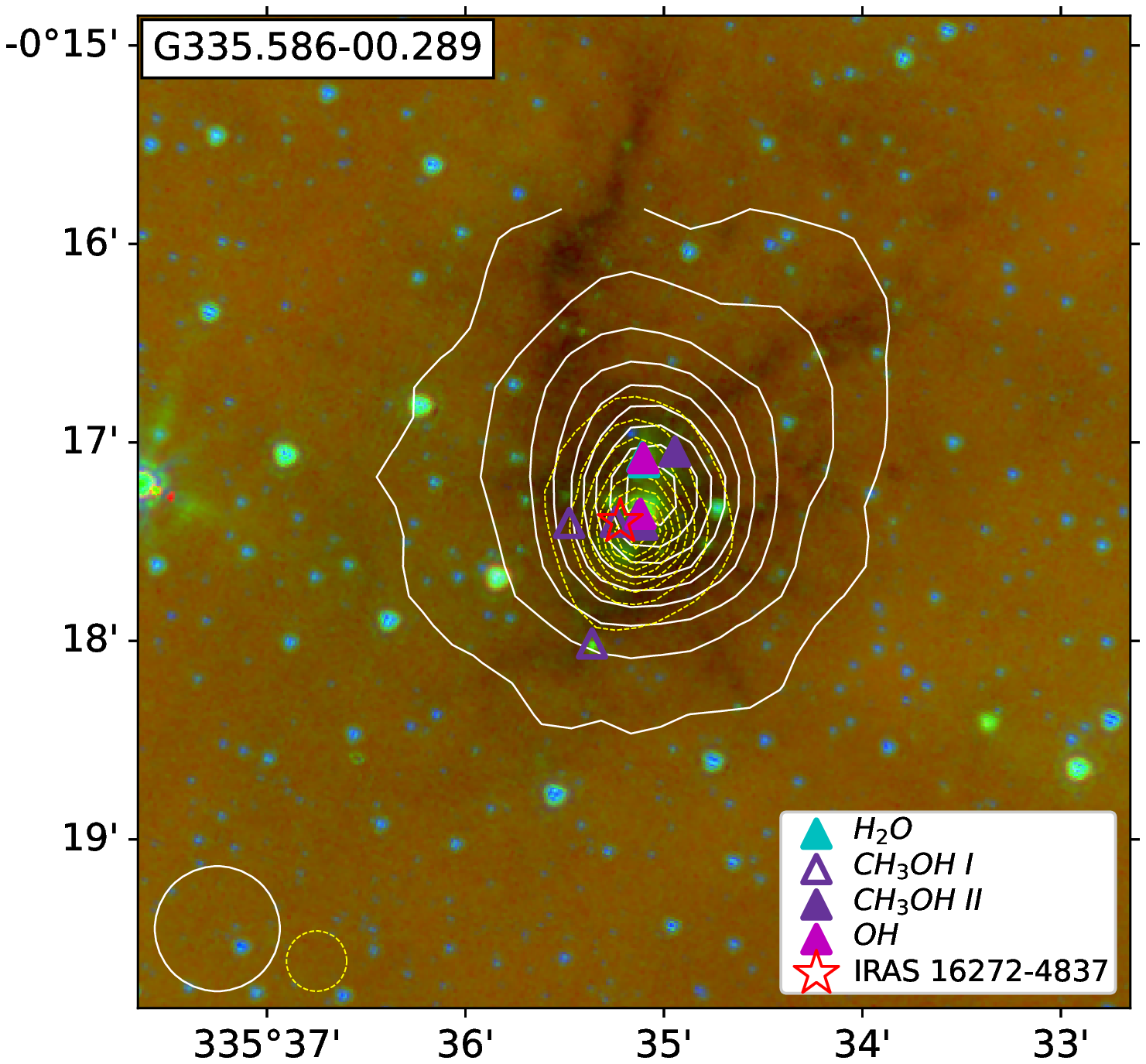}
\end{minipage}

\hspace{5cm}

\begin{minipage}[b]{0.38\textwidth}
    \includegraphics[width=\textwidth,angle=-0]{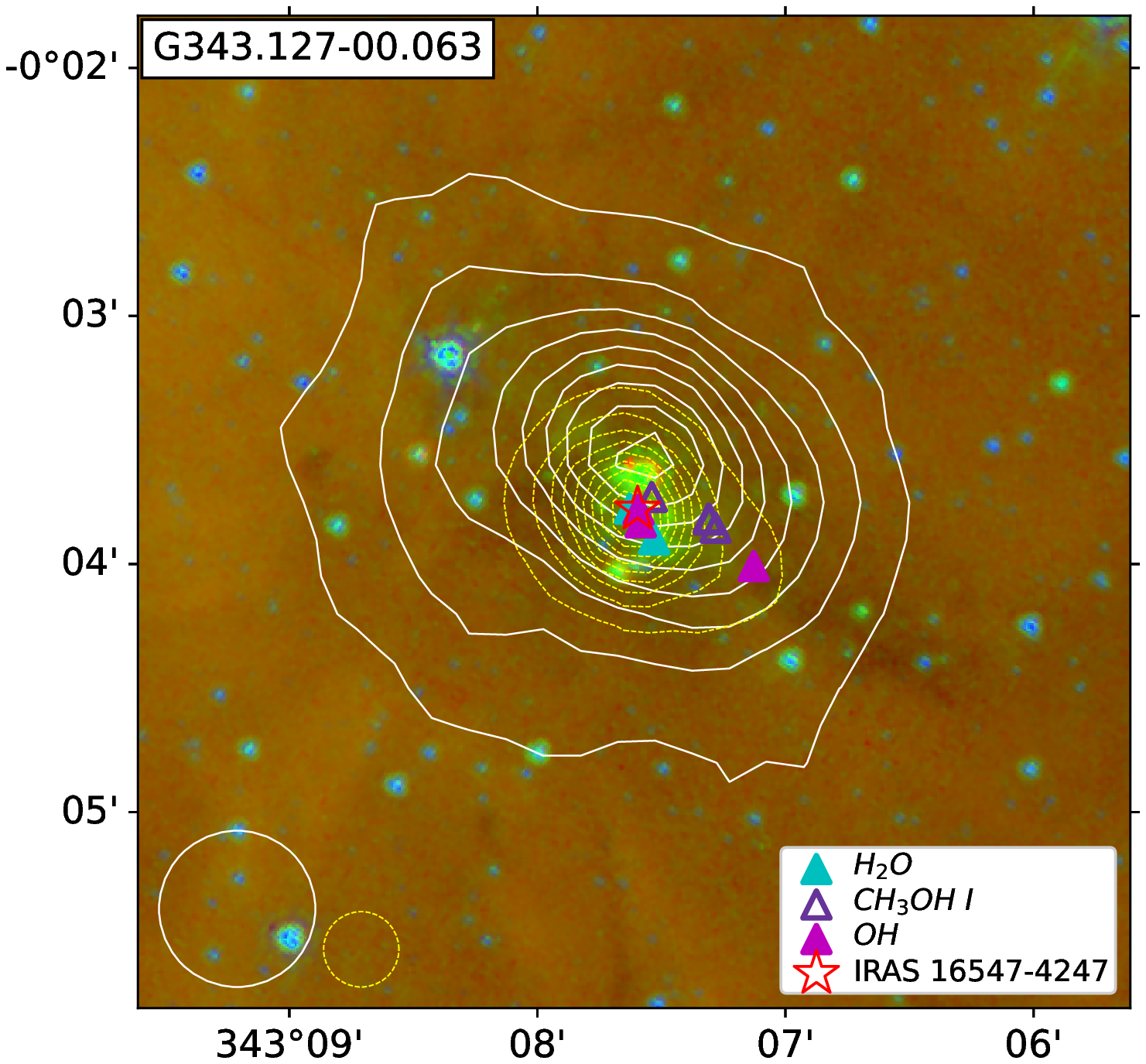}
\end{minipage}

\caption{\scriptsize
Maps of the MALT90 cores in Galactic coordinates. 
Spitzer data are shown in colors: 
8~$\mu$m (red), 5.8~$\mu$m (green), and 3.6~$\mu$m (blue) [23]. 
Yellow dashed contours represent the 870~$\mu$m ATLASGAL data [15]. 
The HCO$^+$(1--0) integrated intensity contours are shown in white. 
The contours vary from 10\% to 90\% of the peak values
which are equal to 7.3, 3.4, 12.5, 12.1, and 18.9 Jy/beam (ATLASGAL) and 
27.8, 21.5, 32.0, 42.8, and 39.0~K\,km/s (HCO$^+$) 
for G012.418, G326.472, G328.567, G335.586, and G343.127, respectively. 
The IRAS and maser source positions as well as 
the main beams of MOPRA-22m (36$''$ at 86~GHz) 
and APEX-12m (19$''$ at 345~GHz) are also shown.
}
\label{mom0}
\end{figure}

\section{Description of objects}

The dense core of G012.418+00.506 is associated
with the source IRAS 18079--1756, which is likely a
massive young stellar object. 
The core has an outflow
[24] and is classified as an extended green object (EGO) [25]. 
It is associated with an ultracompact H II
region [26] and appears to be in a state of global collapse [27]. 
This core was included in the 6 cm continuum
[28] and molecular line [29, 30] surveys, as well
as the ATOMS survey conducted in continuum and
molecular lines with the ALMA interferometer [27].
The core contains water masers [30--32], as well as
methanol masers of classes I [33, 34] and II [35].
\footnote{Information on the presence of maser sources is taken 
from the maserdb.net database [36].} 
The mass of the core according to ATLASGAL data is
414~$M_{\odot}$ [37], the temperature of the dust is 25.6~K [21].

The object G326.472+00.888 is a core located near
the boundary of the expanding H II region, which was
probably formed as a result of compression by a shock wave. 
The center of the 843~MHz radio emission
region [38], as well as the source IRAS 15384--5348,
are located at a distance of pc from the center of the core [39]. 
In addition to ATLASGAL and
MALT90 observations, this core was observed in the
continuum at 350~$\mu$m [40]. 
We did not find data on the
existence of internal sources associated with the core,
which may indicate the earliest stage of evolution
among the objects in the sample. 
Observations of
maser lines in this core were not carried out. 
In [21], this object is classified, however, as ``PDR+Embedded Source''. 
The mass of the core according to
ATLASGAL data is $\sim 280$~$M_{\odot}$ [40], the temperature of
the dust in the core is $\sim 20$~K [21, 39, 40].

The core of G328.567--00.535 is associated with an
extended H~II region [41] and a large-scale bipolar
outflow observed in the near and mid-infrared ranges [42]. 
The core is associated with a cluster of nearinfrared
sources [42] and the source IRAS 15557--5337. 
The evolutionary status according to the classification
of the work [21] is ``complicated''. 
In addition to ATLASGAL, the core was included in a survey of
massive protostars from the southern hemisphere in
molecular lines and in the continuum at 1.2~mm [43].
Observations in the methanol [44] and hydroxyl [45]
lines in this core did not give a positive result. 
The mass of the core according to ATLASGAL data 
is $\sim 470$~$M_{\odot}$ [37]. 
The temperature of the dust in the
direction of the core is $\sim 35$~K [21].

The object G335.586--00.289 is a dark infrared
cloud [46] with a hub-filament system, in which,
according to the conclusions of [47], a global collapse
process is taking place. 
The IRAS 16272--4837 source is associated with the core. 
The core was included in a sample of hot cores observed 
in the continuum at 1.3~mm [48], as well as at 3~mm and 0.87~mm 
and in the H$^{13}$CO$^+$(1--0) line with the ALMA interferometer [47]. 
The evolutionary status according to the classification
of [21] is ``YSO''. According to the ATLASGAL
observations [37], the mass of the core is $\sim 2000$~$M_{\odot}$
for a distance of 3.8~kpc, which corresponds to $\sim 1700$~$M_{\odot}$
for the distance of 3.2~kpc adopted by us. 
According to the data of [49], however, the mass of the core 
is significantly higher and is 3.7$\times 10^3$~$M_{\odot}$. 
The core contains water [50, 51], hydroxyl [52], and methanol
masers of class I [53--55] and class II [56, 57]. The
temperature of the dust in the core is $\sim 23$~K [21].

The core of G343.127--00.063 contains the bright
source IRAS 16547--4247 associated with a massive
O-type star [58]. The evolutionary status of the core
according to the classification of [21] is ``H~II region''.
Interferometric observations in the radio continuum
made it possible to detect a thermal jet coming from
the center of the core in two opposite directions [59, 60]. 
According to observations in the CO(3--2) line
[58], a collimated bipolar outflow oriented along the
radio jet exists in the core. 
Observations with the
ALMA interferometer [61] indicate the existence of a
molecular Keplerian disk around the young massive
stellar object, the orientation of which is perpendicular
to the jet. Maser lines of water [51, 62], hydroxyl [52, 63], 
and methanol class I [53, 54, 64] are observed in
the core. The mass of the core according to
ATLASGAL data is $\sim 890$~$M_{\odot}$ [37], according to
observations at 1.2~mm the mass is $\sim 1.3\times 10^3$~$M_{\odot}$[59].
The temperature of the dust in the core is $\sim 29$~K [21].

The bolometric luminosities of the IRAS sources
associated with four cores (except for
G326.472+00.888) and calculated as the integral of
the fitted ``gray'' body radiation curve into the
frequency dependence of the flux [65] taking into
account the distances from Table~1 are 
(0.1--2)$\times 10^5$~$L_{\odot}$. 
According to our estimates, the source IRAS
15557--5357 has the highest luminosity, and IRAS
18079--1756 has the lowest luminosity.

According to MALT90 data, the following lines
were detected in all cores: HCO$^+$(1--0), N$_2$H$^+$(1--0),
HCN(1--0), HNC(1--0), H$^{13}$CO$^+$(1--0), 
HC$_3$N(10--9), and C$_2$H(1--0, 3/2--1/2, 
F=2--1; 3/2--1/2 F=1--0; 1/2--1/2 F=1--1). 
Weak emission was also
recorded in the lines HN$^{13}$C(1--0) and $^{13}$CS(2--1)
(G328.567), HNCO(4(0,4)--3(0,3)) (G335.586) and SiO(1--0) (G343.127).
\footnote{From here on, abbreviated names of objects are used.}
The emission regions in the continuum
at 870~$\mu$m are close in shape to spherically symmetric (Fig.~1).

To estimate the sizes of the molecular emission
regions, a convolution of a two-dimensional Gaussian
elliptic function with unknown parameters and a twodimensional
circular Gaussian with a width equal to
the width of the main beam of the telescope radiation
pattern was fitted into the integrated intensity maps [66]. 
In the core of G012.418, the size of the HCO$^+$(1--0) 
emission region (FWHM), according to our
estimates, is $\sim 0.5$~pc, and the size of the N$_2$H$^+$(1--0)
emission region is $\sim 0.4$~pc. 
In the core of G326.472,
the shapes of the HCO$^+$(1--0) and N$_2$H$^+$(1--0) emission
regions are close to spherically symmetric (Fig.~1),
and their sizes are also close and amount to $\sim 0.8$~pc.
The shape of the HCO$^+$(1--0) emission region in the
core of G328.567 is more elongated compared to the
other cores (Fig.~1) (the ratio of the axes of the fitted
ellipse is $\sim 2$) (Fig. 1). 
The continuum emission region
at 870~$\mu$m, however, is less elongated (the axial ratio is $\sim 1.4$ [37]). 
The N$_2$H$^+$(1--0) emission peak is shifted
relative to HCO$^+$(1--0) by $\sim 0.5$~pc. This fact, as well
as significant differences in the sizes of the lines HCO$^+$(1--0) 
($\sim 0.8$~pc) and N$_2$H$^+$(1--0) ($\sim 0.2$~pc), indicate
chemical differentiation. The shapes of the HCO$^+$(1--0) 
and N$_2$H$^+$(1--0) emission regions in G335.586 and
in G343.127 are close to spherically symmetric (Fig.~1). 
The sizes of these regions in G335.586 are
$\sim 0.9$~pc and $\sim 0.8$~pc for HCO$^+$(1--0) 
and N$_2$H$^+$(1--0), respectively, and pc for both lines in G343.127.
Figure~2 shows the spectra of HCO$^+$(1--0), HCN(1--0), 
H$^{13}$CO$^+$(1--0) and N$_2$H$^+$(1--0) in the direction
of the integrated intensity peaks of HCO$^+$(1--0).

\begin{figure}[t!]
\setcaptionmargin{5mm}
\onelinecaptionsfalse
\captionstyle{flushleft}

\begin{minipage}[b]{0.32\textwidth}
    \includegraphics[width=\textwidth,angle=-0]{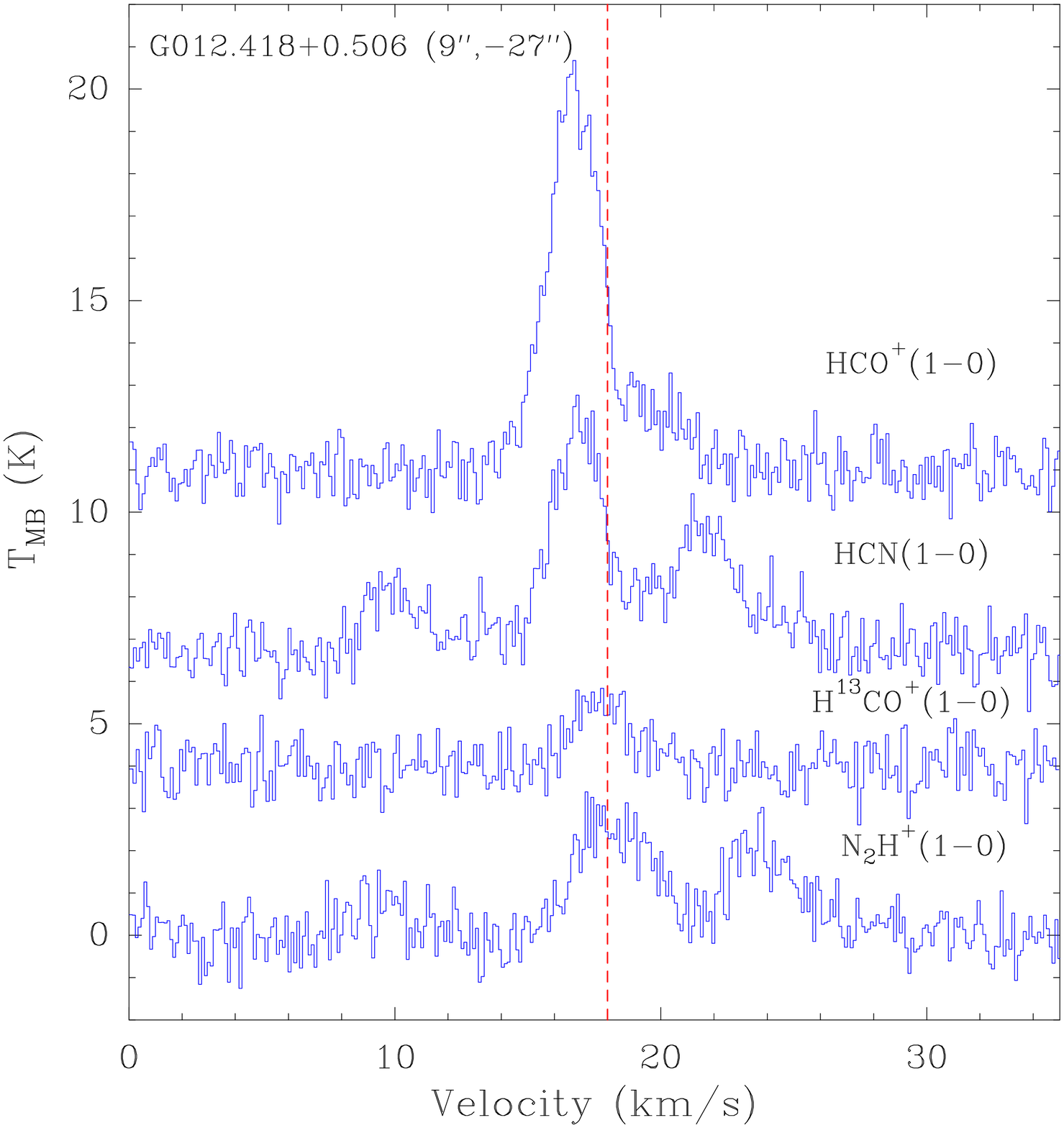}
\end{minipage}
\hspace{1mm}
\begin{minipage}[b]{0.3\textwidth}
    \includegraphics[width=\textwidth,angle=-0]{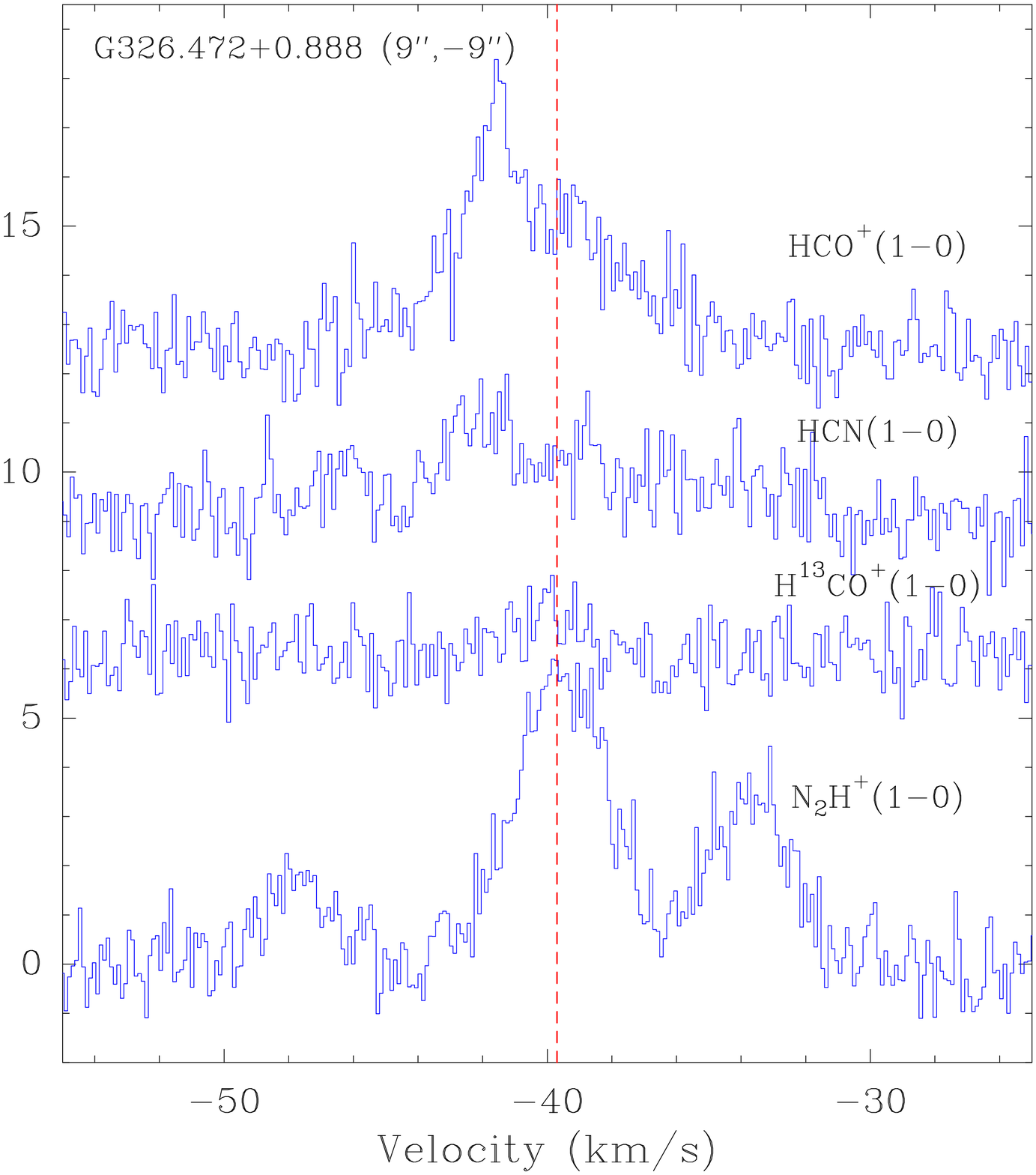}
\end{minipage}
\hspace{1mm}
\begin{minipage}[b]{0.31\textwidth}
    \includegraphics[width=\textwidth,angle=-0]{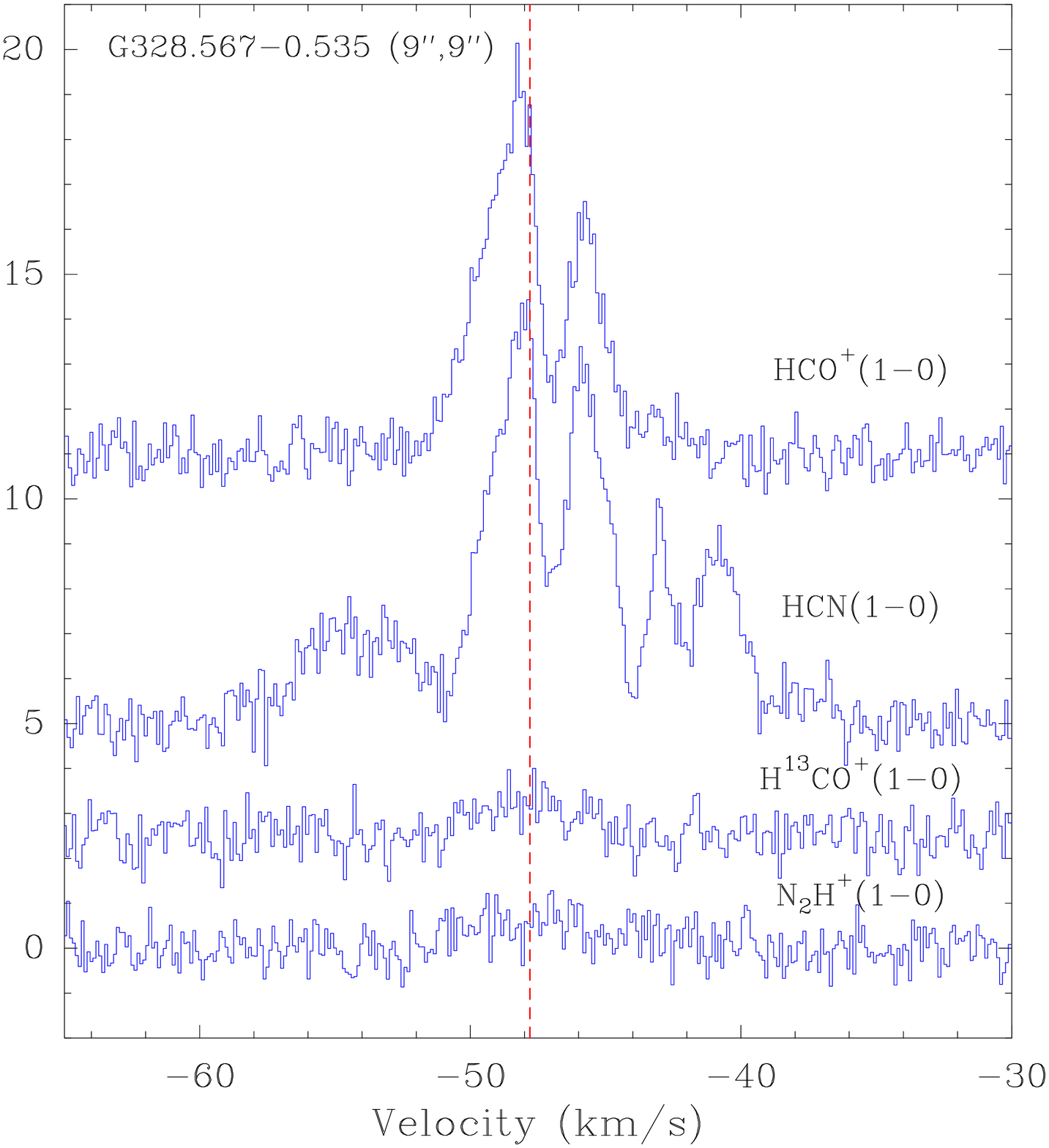}
\end{minipage}

\vspace{3mm}

\begin{minipage}[b]{0.32\textwidth}
    \includegraphics[width=\textwidth,angle=-0]{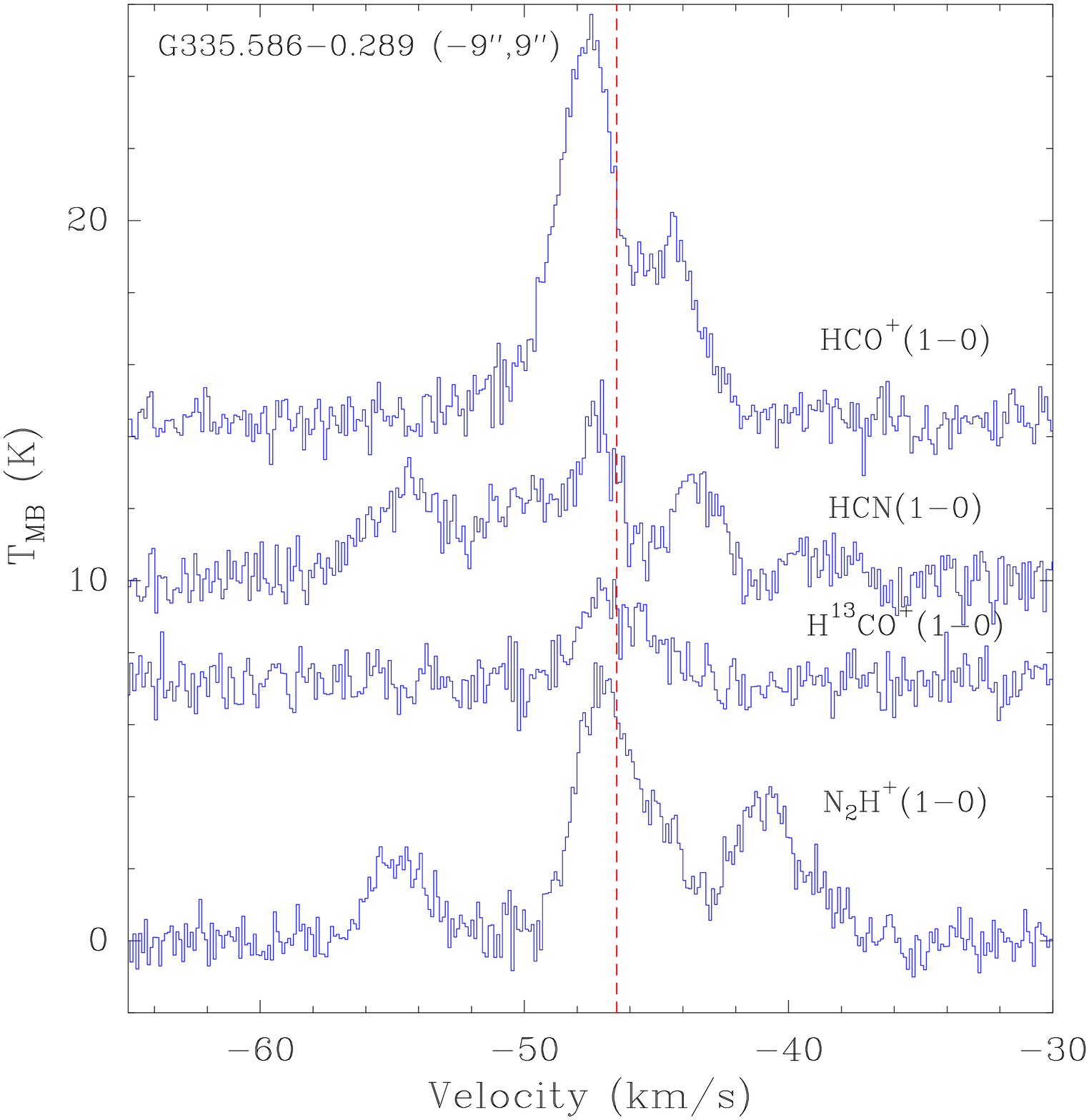}
\end{minipage}
\hspace{1mm}
\begin{minipage}[b]{0.3\textwidth}
    \includegraphics[width=\textwidth,angle=-0]{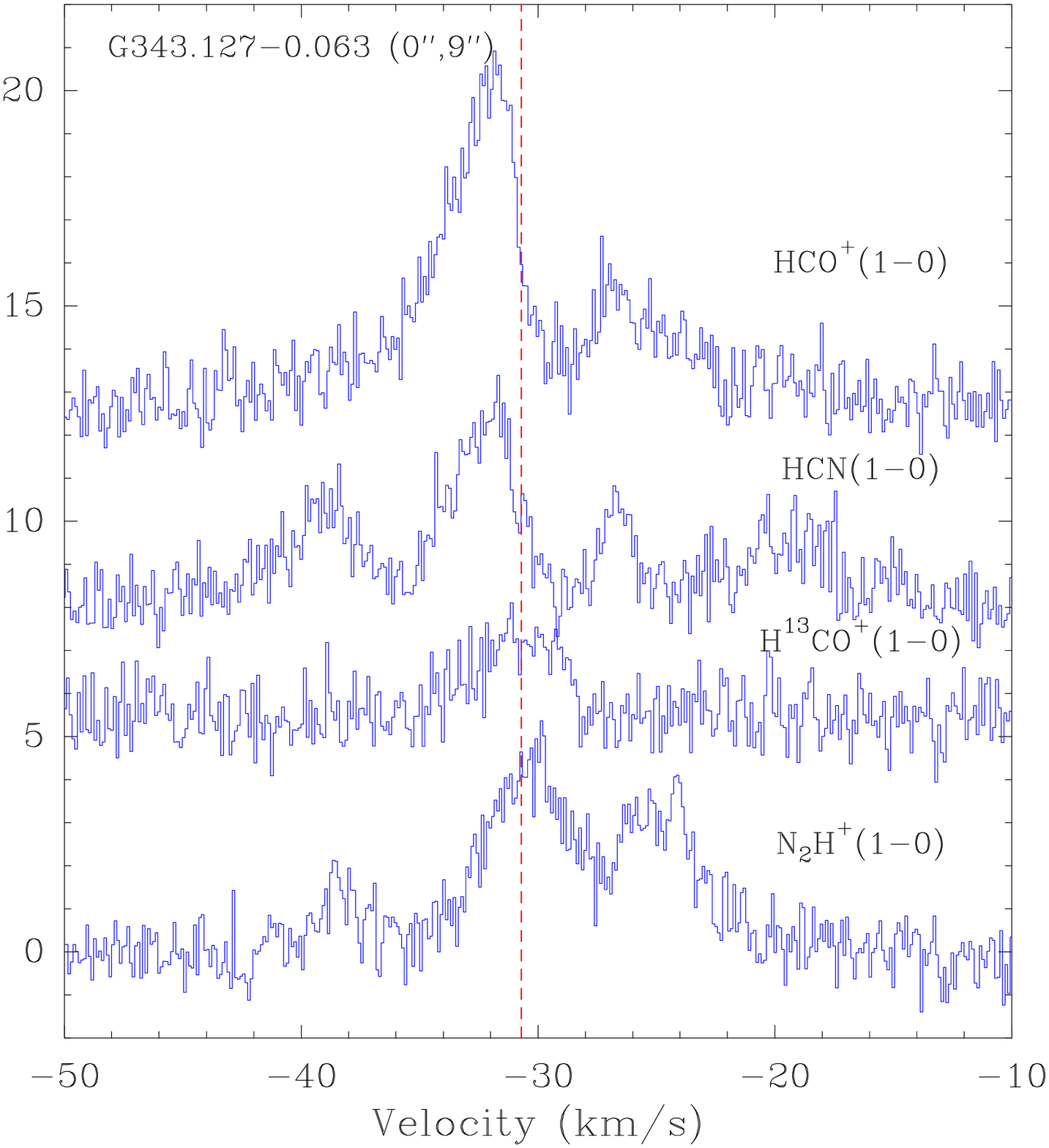}
\end{minipage}

\caption{\scriptsize
Spectra in the direction of emission peaks in the cores from the sample. 
Each figure shows the name of the core and its
position on the map (shift relative to the coordinates given in Table 1). 
The horizontal axes show the radial velocity, and the vertical
axes show the intensity in units of temperature reduced to the main beam 
of the telescope ($T_{MB}$). The vertical dashed red
lines correspond to the centers of optically thin H$^{13}$CO$^+$(1--0) lines, 
which coincide with the centers of the N$_2$H$^+$(1--0) lines 
in all cores except G335.586.
}
\label{fig:spectra}
\end{figure}

\section{Estimation of physical parameters of cores
using model calculations}

To obtain information on the radial profiles of
physical parameters in the cores, including contraction
velocity, we fitted model spectral maps into the
observed ones. 
As in [9, 13], we used a multilayer
spherically symmetric model of the core (SSL), the
parameters of which (density, kinetic temperature,
turbulent and systematic velocities) depended only on
the radial distance, as $P_0/(1+(r/R_0)^{\alpha_p})$
(the description of the model is given in the Appendix 
to the paper [9]). 
The quantity $P_0$ is twice the value of the parameter
in the central layer with the radius , which was
set equal to $5\times 10^{-2}$~pc. 
The parameters of the model
were the values of for the radial profiles of the density,
turbulent and systematic velocities ($n_0$, $V_{turb}$, $V_{sys}$
respectively), and the corresponding power indices
($\alpha_n$, $\alpha_{turb}$, $\alpha_{sys}$), as well as 
the relative abundances of the HCO$^+$ and H$^{13}$CO$^+$ molecules, 
which were assumed to be independent of the radial distance. 
The latter condition can be a fairly rough approximation
for the cores of the sample. It is known that even in
low-mass prestellar cores, chemical modeling predicts
a non-uniform distribution of HCO$^+$ [67]. 
However, our calculations using a model in which the HCO$^+$
abundance varied from the center to the edge by several
times did not lead to changes in the estimates of
the core parameters that go beyond the confidence ranges. 
It is likely that when analyzing higher-quality
observational data with better angular resolution, taking
into account the results of chemical model calculations
may be of significant importance.

The kinetic temperature profile for the four cores
with internal sources was specified as
$60~K/(1+(r/R_0)^{0.3}$) and was not varied.
This type of profile corresponds to the dust temperature
distribution in the optically thin dust shell surrounding
the central source for the dust emissivity
value $\beta=2$ [68, 69] provided that the dust and gas
temperature distributions are close. Note that the
kinetic temperature distribution in the cores (especially
at the periphery) can be more complex [70, 71].
However, since the HCO$^+$--H$_2$ collision probabilities
we used [72] are specified for fixed temperatures with
a step of 10~K, the temperatures in the layer were
rounded to a value multiple of 10~K, which leveled out
possible temperature variations in the outer layers. 
In accordance with the chosen profile, the kinetic temperature
was 30~K in the central layer and dropped to
K at the periphery. 
For the core of G326.472,
which apparently has no internal heating source, the
kinetic temperature was assumed to be constant at
20~K. After calculating the model spectra, a convolution
with the telescope radiation pattern was performed
and an error function was calculated that
depends on the difference between the model and
observed spectra at different points.

In addition to the SSL code, we used the LOC code
[73], which uses a deterministic set of rays in Cartesian
coordinates to calculate radiative transfer and which is
applicable to an arbitrary distribution of parameters in
the model cloud, allowing, in particular, to take rotation
into account.

The method of applying the algorithm for finding
the global minimum of the error function is based on
the scheme described in [9, 13]. Since the calculations
of the excitation of molecules and the transfer of radiation
in lines obviously do not depend on the choice of
a specific object and the distance to it, a general library
of spectral maps corresponding to a distance to an
object of 1~kpc without convolution with the telescope
radiation pattern was preliminarily calculated. 
When analyzing individual sources, the maps were convolved
with the telescope radiation pattern and the
error function was calculated. 
For the values of the
physical parameters corresponding to the minimum of
the error function, the boundaries of the confidence
ranges were calculated. 
To refine the estimates, the
Nelder--Mead method and the LOC program were
used. In this case, the values calculated from the analysis
of the library of spectral maps were taken as initial values. 
As was shown in [9, 13], the error function has
a single global minimum. 
When choosing initial values
in the vicinity of this minimum, the Nelder--Mead
method showed sufficient efficiency and allowed us to
estimate the optimal values of the physical parameters
when calculating models with a different type of
parameterization (with a fixed turbulent velocity profile
and taking rotation into account, see Section~5).

The observed and model HCO$^+$(1--0) maps of the
central regions of the cores are shown in Fig.~3. 
The values of the physical parameters corresponding to the
minimum of the error function, as well as the uncertainties
of these estimates, corresponding to the
boundaries of the confidence ranges, are given in
Table~2.

\begin{table}[p]
\setcaptionmargin{0mm}
\onelinecaptionsfalse

\centering
\caption{Model values of physical parameters of cores}
\vskip 2mm
\scriptsize
\begin{tabular}{l|rr|rr|rr|rr|rr|}
%\noalign{\hrule}\noalign{\smallskip}
\noalign{\hrule}\noalign{\smallskip}
Parameter & \multicolumn{2}{|c|} {G012.418+00.506}
         & \multicolumn{2}{|c|} {G326.472+00.888}
         & \multicolumn{2}{|c|} {G328.567--00.535}
         & \multicolumn{2}{|c|} {G335.586--00.289}
         & \multicolumn{2}{|c|} {G343.127--00.063} \\
%Параметр & G012.418+00.506 & G326.472+00.888  & G328.567--00.535 & G335.586--00.289 & G343.127--00.063 \\
\noalign{\hrule}\noalign{\smallskip}
$\log(n_0)$               & 7.3$^{+0.1}_{-1.1}$   & 7.2   &  7.3$^{+0.7}_{-0.4}$   &6.7   & 7.9$^{+0.1}_{-0.6}$    &7.4   & 8.0$^{+0.2}_{-1.0}$    &7.8  &  7.0$^{+0.8}_{-0.2}$   &6.4   \\
$\alpha_n$                & 2.3$^{+0.1}_{-0.2}$   & 1.5   &  1.5$^{+0.3}_{-0.1}$   &1.2   & 1.8$^{+0.1}_{-0.5}$    &1.5   & 2.8$^{+0.4}_{-0.8}$    &2.5  &  2.0$^{+0.3}_{-0.2}$   &1.8   \\
V$_{turb}$ (km/s)         & 3.7$^{+1.4}_{-0.2}$   &       &  10.4$^{+3.7}_{-1.1}$  &      & 6.8$^{+6.8}_{-1.4}$    &      & 5.0$^{+2.9}_{-1.8}$    &     &  17.3$^{+0.2}_{-5.8}$  &      \\
$\alpha_{turb}$           & 0.3$^{+0.02}_{-0.1}$  &       &  0.5$^{+0.1}_{-0.08}$  &      & 0.35$^{+0.23}_{-0.08}$ &      & 0.30$^{+0.2}_{-0.05}$  &     &  0.6$^{+0.01}_{-0.14}$ &      \\
V$_{sys}$(km/s)           & --2.5$^{+0.9}_{-0.7}$ &--1.9  &--2.3$^{+0.5}_{-0.4}$   &--5.5 &--1.1$^{+0.1}_{-1.1}$   &--3.5 &--2.1$^{+0.7}_{-0.6}$   &--2.7&  --3.0$^{+0.3}_{-2.5}$ &--3.2 \\
$\alpha_{sys}$            & 0.12$^{+0.06}_{-0.06}$&  0.0  & 0.13$^{+0.07}_{-0.09}$ &0.09  & 0.06$^{+0.14}_{-0.03}$ &0.0   &  0.11$^{+0.11}_{-0.06}$& 0.1 &  0.07$^{+0.1}_{-0.07}$ &0.09  \\
$\log{X}$(HCO$^+$)        & --10.2$^{+0.3}_{-0.7}$&--10.7 & --10.9$^{+0.4}_{-0.4}$ &--10.8& --10.8$^{+0.3}_{-1.1}$ &--10.8& --9.6$^{+0.2}_{-0.6}$  &-10.0& --9.6$^{+0.3}_{-0.7}$  &--9.2 \\
\noalign{\smallskip}\hline\noalign{\smallskip}
\end{tabular}
\flushleft
{\small
$n_0$, V$_{turb}$ and V$_{sys}$
represent doubled values of the parameters in the central layer with the radius
$R_0=5\times 10^{-2}$~pc; $n_0$ has the dimension [cm$^{-3}$]. 
On the right in each column for a given object are the results of calculations 
with a fixed turbulent velocity profile (see text).
}
\label{table:modelpar}
\end{table}

\begin{figure}[t!]
\setcaptionmargin{5mm}
\onelinecaptionsfalse
\captionstyle{flushleft}

\begin{minipage}[b]{0.8\textwidth}
    \includegraphics[width=\textwidth,angle=-0]{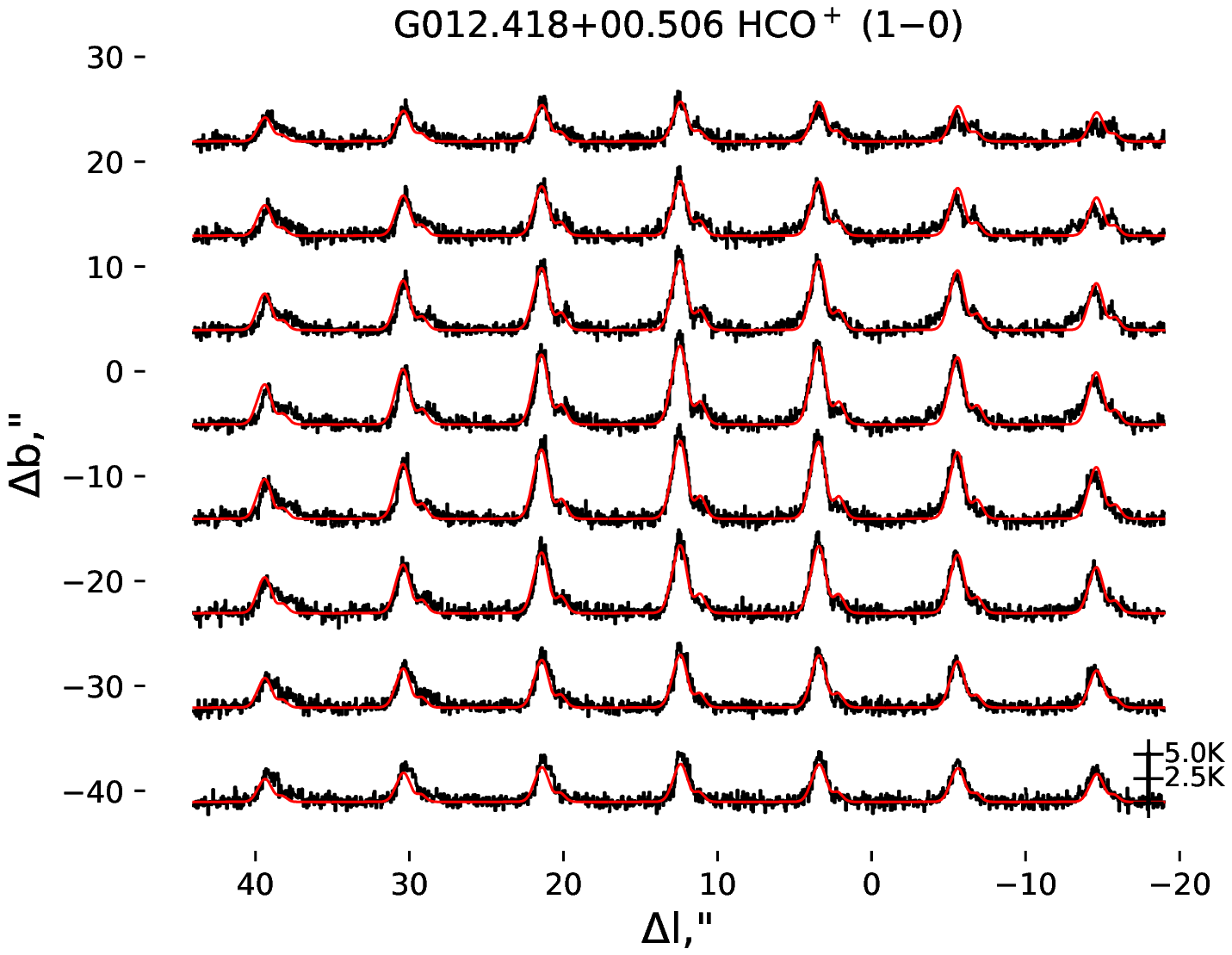}
\end{minipage}

\vskip 3mm

\begin{minipage}[b]{0.8\textwidth}
    \includegraphics[width=\textwidth,angle=-0]{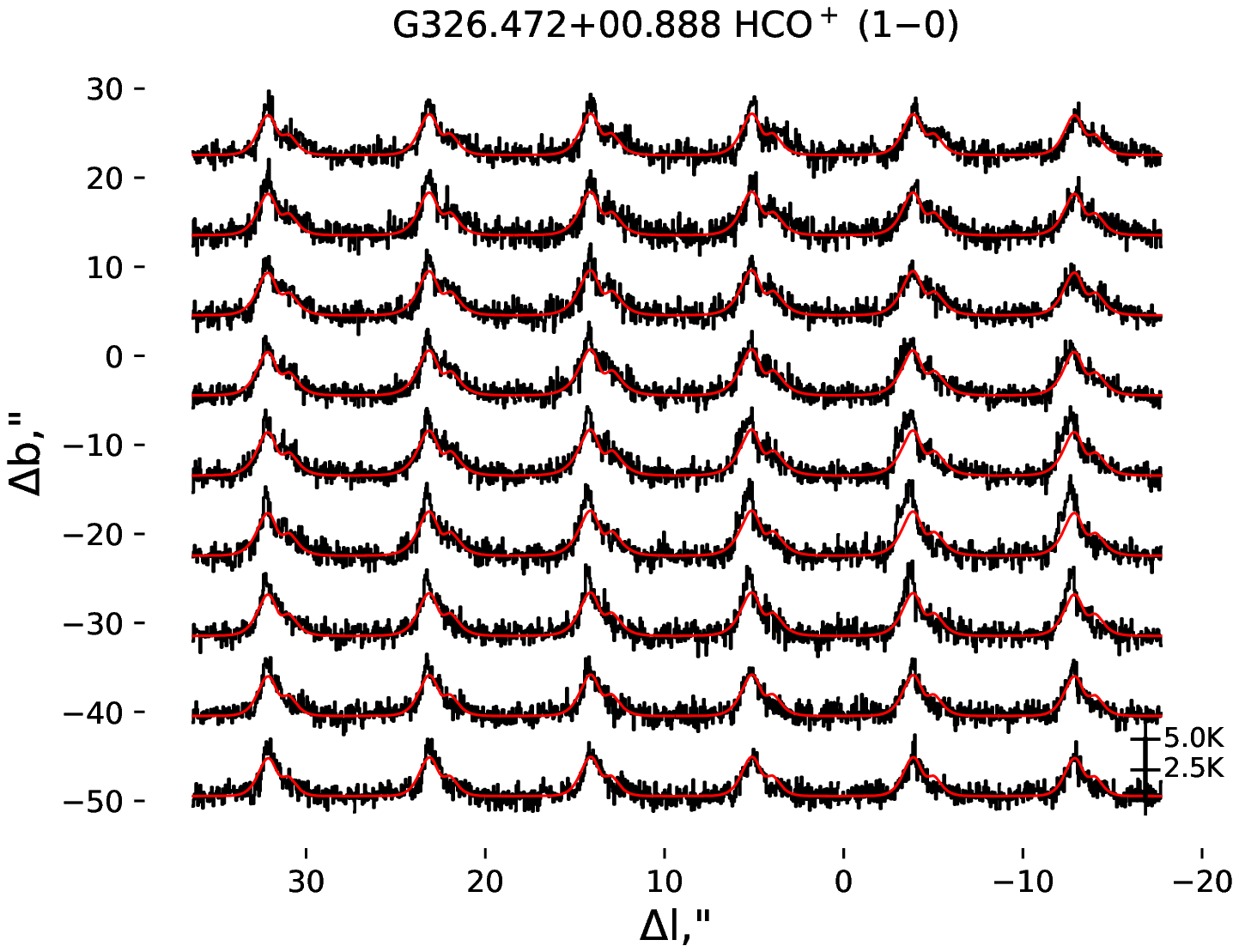}
\end{minipage}

\caption{\scriptsize
Observed (black) and model (red) HCO$^+$(1--0) maps of the cores. 
The axes are the offsets relative to the coordinates given in Table~1. 
The intensity scale is indicated in the lower right corner of each figure. 
Velocity varies from 7 to 26~km/s, from --52 to --29~km/s,
from --65 to --30~km/s, from --54 to --39~km/s, and from --42 to --20 km/s 
for G012.418, G326.472, G328.567; G335.586 and G343.127, respectively.
}
\label{fig:model}
\end{figure}

\addtocounter{figure}{-1}

\begin{figure}[t!]
\setcaptionmargin{5mm}
\onelinecaptionsfalse
\captionstyle{flushleft}

\begin{minipage}[b]{0.9\textwidth}
    \includegraphics[width=\textwidth,angle=-0]{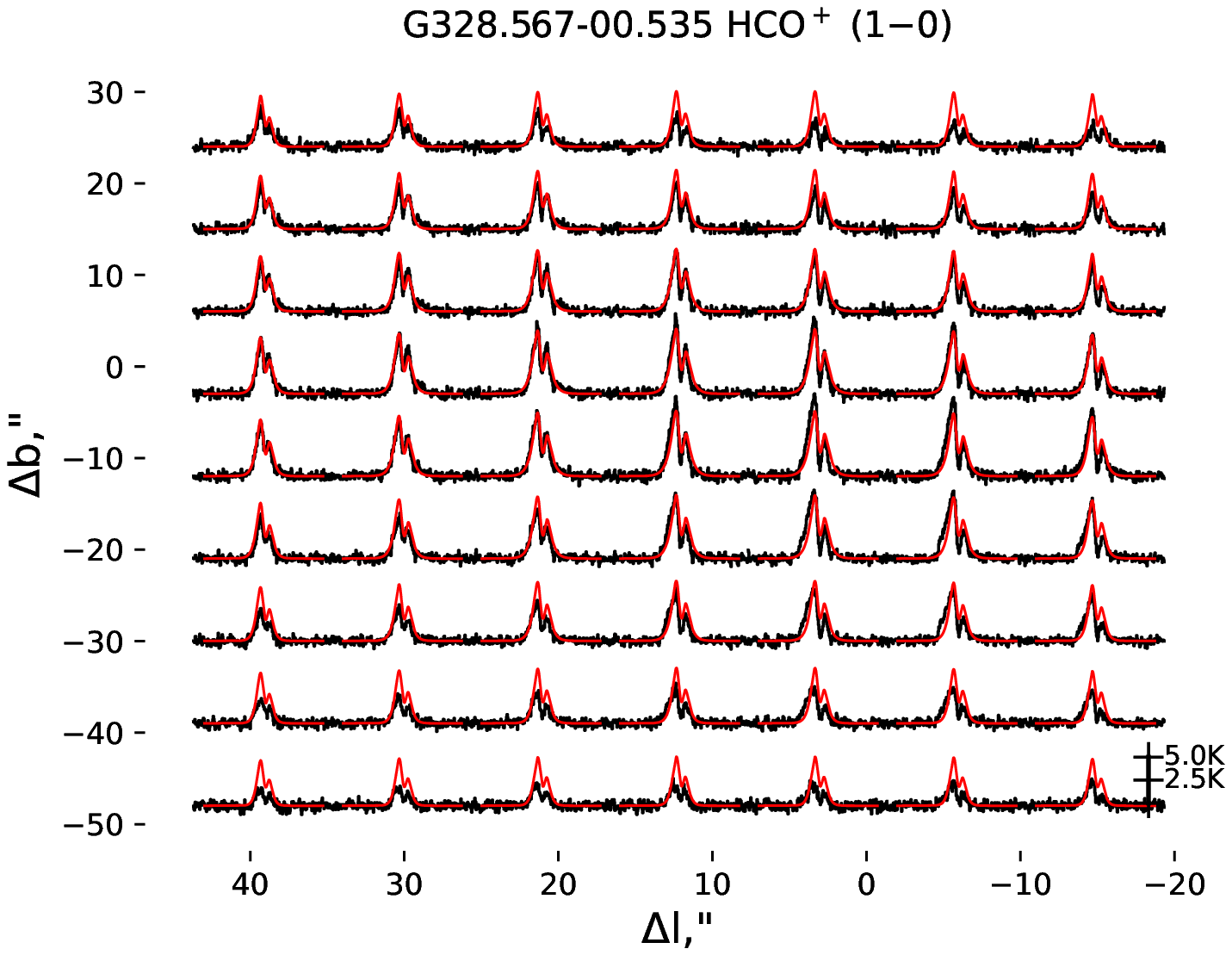}
\end{minipage}

\vspace{3mm}

\begin{minipage}[b]{0.9\textwidth}
    \includegraphics[width=\textwidth,angle=-0]{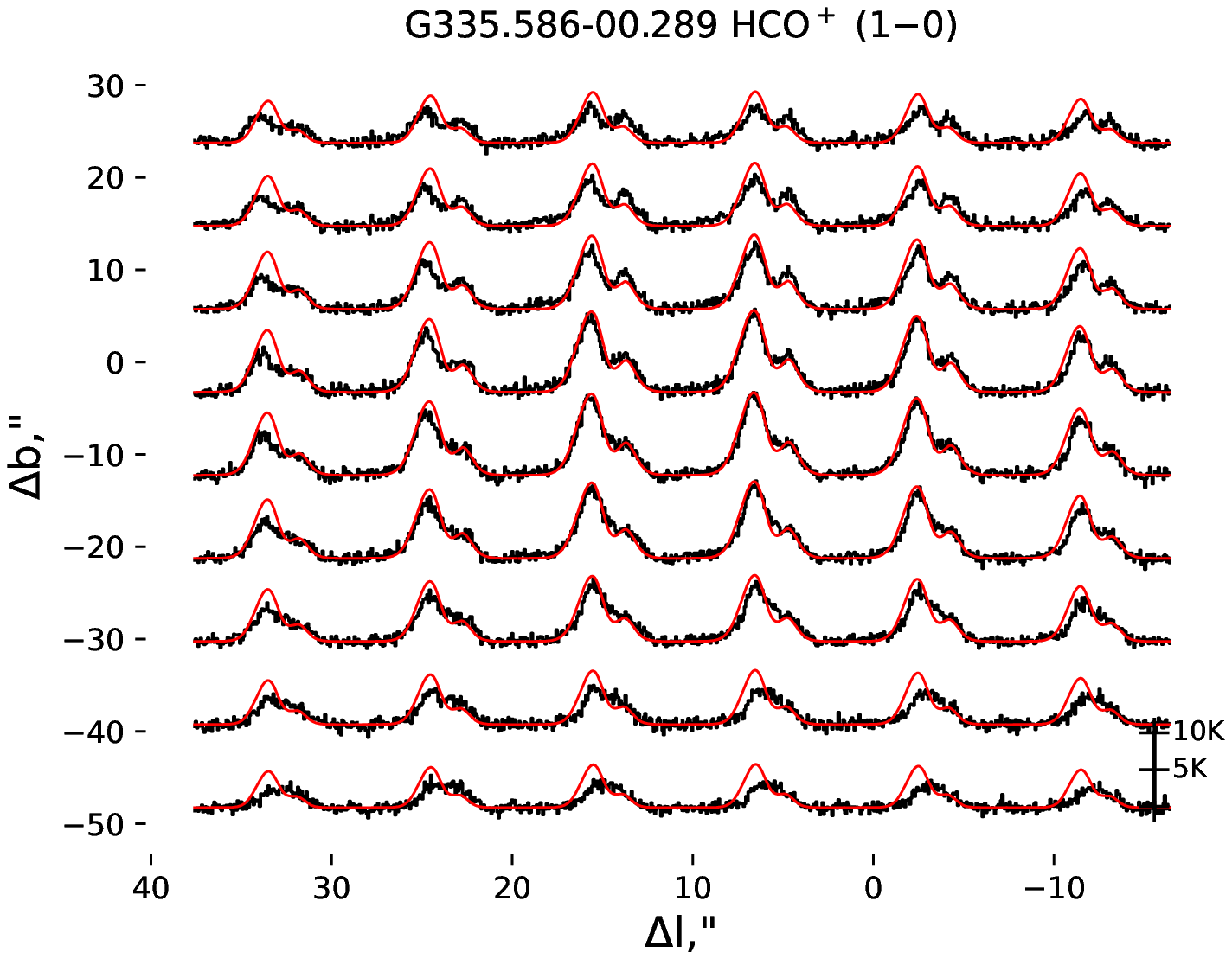}
\end{minipage}

\caption{\scriptsize
Continued}
\end{figure}

\addtocounter{figure}{-1}

\begin{figure}[t!]
\setcaptionmargin{5mm}
\onelinecaptionsfalse
\captionstyle{flushleft}

\begin{minipage}[b]{0.9\textwidth}
    \includegraphics[width=\textwidth,angle=-0]{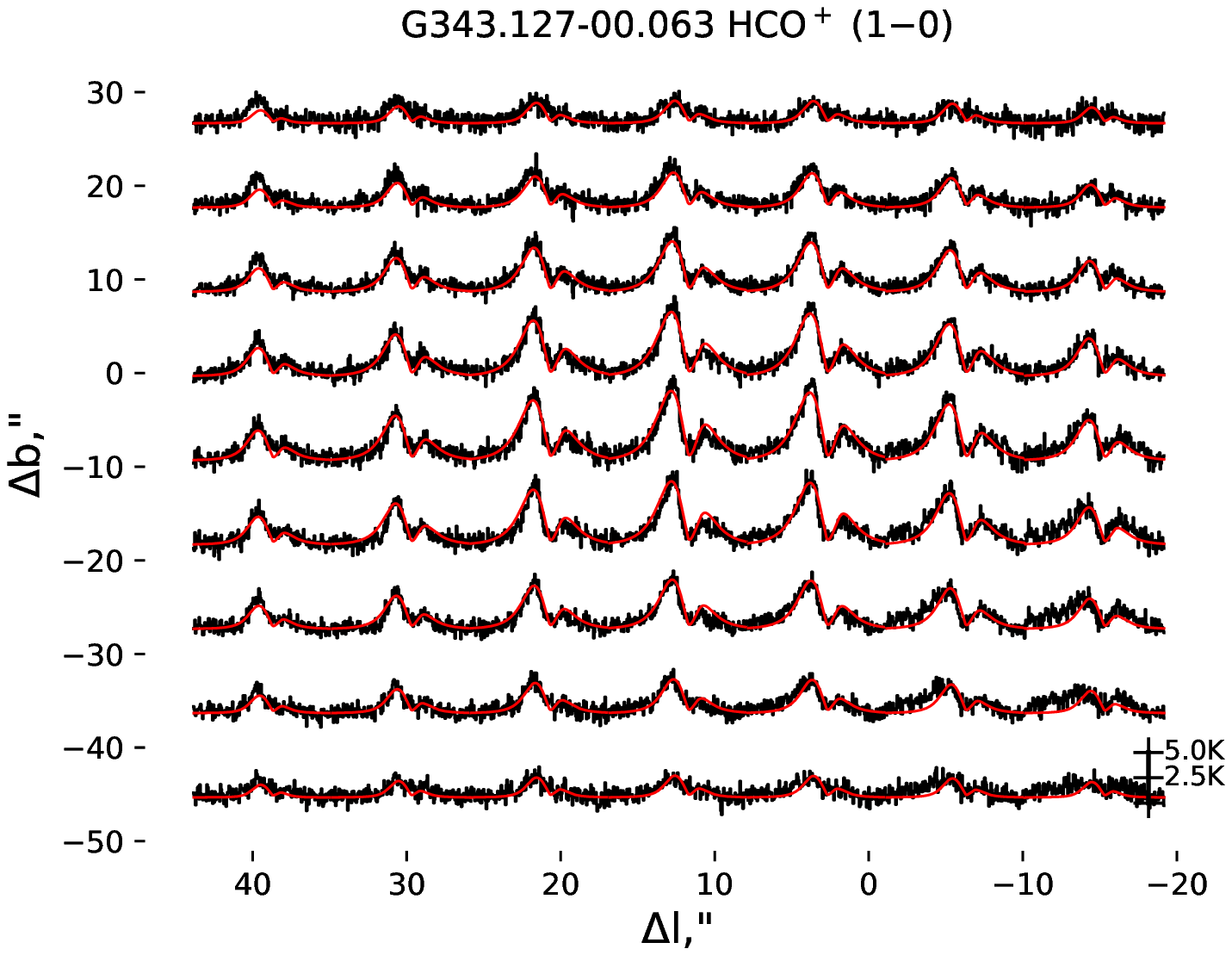}
\end{minipage}

\caption{\scriptsize
Continued
}
\label{fig:model}
\end{figure}

\section{Discussion}
\label{sec:discussion}

The dense MALT90 cores included in the sample
represent regions of formation of massive stars at different
stages of evolution. 
The earliest stage of evolution
is probably represented by the core G326.472,
where there are no indications of the presence of internal sources. 
However, the widths of optically thin lines
in the central regions of this core are higher than at the
periphery, indicating an increased degree of dynamical
activity in the center (turbulence, systematic
motions), which can be associated with the process of star formation. 
The other four cores are apparently at
later stages, as indicated by the presence of high-luminosity
IRAS sources, H~II zones, maser sources and
other sources (see Section 3). 
Among them, the core
of G328.567 stands out, associated with an extended
H~II zone, a large-scale bipolar outflow and a cluster
of infrared sources, where there are no data on the
presence of maser sources. The center of the N$_2$H$^+$(1--0) emission 
region in this core is shifted relative to
the HCO$^+$(1--0) center, indicating chemical differentiation,
probably associated with the influence of a
massive young stellar object. 
For the remaining cores,
the morphologies of the HCO$^+$(1--0) and N$_2$H$^+$(1--0)
emission regions, as well as the sizes of their emission regions, are close. 
The sizes of the emission regions in
the HCO$^+$(1--0) line vary in the range of $0.4-1.0$~pc (Section~3). 
The most compact core is G012.418. 
The dust temperatures are in the range of $\sim 20-35$~K [21].
The core of G328.567 has the highest dust temperature.
The masses of the cores calculated from the dust
emission in the continuum are in the range from $\sim 300$
to $\ge 1000~M_{\odot}$ [37, 40]. 
The core of G335.586 has the largest mass. 
The virial masses calculated from
the H$^{13}$CO$^+$(1--0) lines (for G328.567) and N$_2$H$^+$(1--0) 
(for the remaining cores) are close to the mass estimates
calculated from dust, within possible errors, the
main contribution to which may apparently be associated
with uncertainties in the kinematic distances.

\subsection{Analysis of the Results of Model Calculations}

By fitting the model maps into the observed maps
of HCO$^+$(1--0) and H$^{13}$CO$^+$(1--0), the parameters of
the radial dependences of the density, turbulent velocity,
and contraction velocity of the cores were estimated (Table~2). 
The results of the model calculations
describe the observed spectra quite well (Fig.~3). 
The boundaries of the confidence regions for the parameter
values in the center were in some cases strongly
asymmetric, which is apparently due to correlations
between the parameters and the nonlinear dependence
of the error function on the parameters.

The power indices of density decay with distance
from the center ($\alpha_n$) lie in the range $\sim 1.5-2.8$. 
The lowest index was obtained for the core of G326.472,
the highest -- for the core G335.586. 
The range of confidence
estimates in the latter case, however, is quite
large (+0.4/--0.8). 
The calculated values within the
confidence estimates are comparable with the value
$1.6\pm 0.3$
obtained from observations in the continuum
for cores associated with regions of formation of massive
stars and star clusters at different stages of evolution [74]. 
For four cores with internal sources, the values
of the indices $\alpha_n$, taking into account the confidence
ranges, are close to the value of 2 predicted by theoretical models.

The model masses of the cores calculated from the
values of the densities at the center ($n_0$) and the indices
$\alpha_n$ (Table 2) turned out, however, to be higher than the
estimates calculated from the dust observation data
[37, 40], as well as those calculated by us from the
C$^{18}$O(2--1) observation data [17]. 
The strongest discrepancy was noted for the cores of G012.418,
G326.472 and G328.567 ($\ge$ order of magnitude). 
For the cores of G335.586 and G343.127, the discrepancies
do not go beyond the range of confidence estimates,
taking into account the large uncertainties of
the $n_0$ values (Table~2). Also, as a result of the calculations,
the turbulent velocity at the center of the cores
turns out to be quite high. The highest value was
obtained for G343.127, which may be associated with
the presence of molecular outflow in this core [58],
which is the cause of additional broadening of the
HCO$^+$(1--0) lines in the center. With increasing distance
from the center, the turbulent velocity drops
quite sharply with indices of $\sim 0.3-0.6$, reaching values
of $\le 1$~km/s in the outer layers.

The systematic excess of model masses over independent
estimates may indicate a common cause
unrelated to errors in parameter estimates. 
The factor leading to overestimated central densities and, 
accordingly, masses may be insufficient spatial resolution 
of observations and probable anisotropic structure of the
density and velocity field, unresolved by observations.
As shown by interferometric observations of G012.418
and G335.586 [27, 47], the central regions of these
cores are highly inhomogeneous. They consist of fragments
with different velocities and different line asymmetries.
In addition, gas and dust in these regions are
concentrated in filaments along which gas motions
apparently occur. MALT90 observational data with
low angular resolution in this case provide only averaged
information on the radiation coming from these
regions, and the use of a spherically symmetric model
may lead to a shift in the estimates of the physical
parameter values. To a greater extent, these effects can
affect the values of the parameters in the center ($P_0$)
and to a lesser extent, the values of the power indices
($\alpha_p$), determined by the observed spectral maps of the
entire core.

To check how the value of the central layer radius
affects the obtained results, we performed calculations
for $R_0=6.5\times 10^{-3}$~pc. A decrease in $R_0$ led to a significant
increase in the density in the center (by $\sim 2$ orders
of magnitude), to an increase in the turbulent velocity
in the center (by $\sim 1.5-2$ times) and to a lesser extent
led to a change in the values of the systematic velocity
and power indices. In particular, the $\alpha_{sys}$ values
remained practically unchanged. A check was also
performed to what extent a possible overestimation of
the turbulent velocity in the center affects the bias in
the estimates of other parameters. For this purpose,
fixed turbulent velocity profiles calculated from observations
of lines with a small optical depth (N$_2$H$^+$(1--0) 
and/or H$^{13}$CO$^+$(1--0)) were taken for each core.
These turbulent velocity profiles had a shape different
from that used in the model calculations, changing
slightly in the central layers and decreasing to $\le 1$~km/s
in the outer layers. 
In this case, the Nelder--Mead
method and the LOC program [73] were used to calculate
the values of the remaining parameters. 
As a result, the central densities $n_0$ and indices $\alpha_n$
decreased, leading to a decrease in the mass estimates;
the values of $V_{sys}$ increased in absolute value, and the
values of $\alpha_{sys}$ remained practically unchanged. 
In Table~2, the values obtained for the case of a fixed turbulent
velocity profile are given in each column on the
right for comparison.

The contraction velocities in the central layer lie in
the range from approximately --0.6 to --1.5~km/s.
These values are close to the estimates calculated using
the model [8] for the centers of the cores of G328.567,
G335.586, and G343.127 based on the parameters of
the observed HCO$^+$(1--0) lines, where the dip in the
lines is quite pronounced.

\subsection{Estimates of the Rotation of the Cores}

In some areas of the maps (see Fig.~3), the ratio
between the ``blue'' and ``red'' peaks of the model
spectra differs from the observed ones, which may be
caused by rotation. The LOC program was used to
estimate the rotation parameters in the cores of
G328.567 and G335.586. For the core of G328.567,
the HCO$^+$(1--0) data were used, and for the core of
G335.586, the archived data of APEX-12m observations
in the HCO$^+$(3--2) line (project C-092.F-9702B-2013) were used, 
which have a better signal-to-noise
ratio than the MALT90 data. The rotation velocity
was specified as $V_{rot}(R_0/r)^{\alpha_v}$. 
The rotation axis was
taken to lie in the plane of the sky in the direction of
the minimum of the gradient of the first moment map
of HCO$^+$(1--0). When fitting the model maps to the
observed ones, $n_0$, $V_{sys}$, $\alpha_n$, and $\alpha_{sys}$ 
were chosen as free parameters, as well as the rotation parameters 
($V_{rot}$ and $\alpha_v$) and the abundance of HCO$^+$ molecules. 
The turbulent velocity profile was assumed to be fixed. 
The optimal rotation parameters were calculated using the
Nelder--Mead method and are given in Table~3. 
The resulting spectral maps are shown in Fig.~4. 
As a result of the calculations, the error function value decreased
by 20\% for G328.567 and by 50\% for G335.586. 
For the core of G328.567, the index $\alpha_v$ turned out to be
close to 0.4, which may indicate a rotation law close to
the Keplerian one. Taking into account the rotation in
this core led to a significant decrease in the estimate of $V_{sys}$
(cf. the corresponding values from Table~2, right
column, and from Table~3). The power index of the
radial contraction velocity profile did not change. 
For G335.586, the rotation velocity depends weakly on the
radial distance.
%indicating a rotation close to a solid state. 
This region may be at an earlier stage of evolution
compared to G328.567, where there is a large-scale
bipolar outflow and, probably, a disk structure.
This conclusion is consistent with the results of model
calculations from [75], which show that the primary
disk at an early stage of evolution can rotate according
to a law close to a solid state. An alternative explanation
for the variations in the line asymmetry in the core
of G335.586 may be the anisotropic nature of accretion
associated with gas flows along the filaments [76].
To obtain more definitive conclusions, further observational
studies of this core should be carried out,
which will allow us to estimate its kinematics as a whole.

\begin{table}[p]
\setcaptionmargin{0mm}
\onelinecaptionsfalse

\centering
\caption{
Model values of physical parameters of cores taking into account rotation
}
\vskip 2mm
\scriptsize
\begin{tabular}{l|l|l|}
\noalign{\hrule}\noalign{\smallskip}
Parameter                 & G328.567--00.535 & G335.586--00.289 \\
\noalign{\hrule}\noalign{\smallskip}
$\log(n_0)$               & 7.8    &  6.7   \\
$\alpha_n$                & 1.9    &  2.2   \\
V$_{sys}$(km/s)           & --0.8  &--2.7   \\
$\alpha_{sys}$            & 0.0    & 0.1    \\
V$_{rot}$(km/s)           & 2.3    & 0.8    \\
$\alpha_{v}$              & 0.4    & 0.1    \\
$\log{X}$(HCO$^+$)        & --10.5 & --9.0  \\
\noalign{\smallskip}\hline\noalign{\smallskip}
\end{tabular}
\flushleft{}
\label{table:rot}
\end{table}

\begin{figure}[t!]
\setcaptionmargin{5mm}
\onelinecaptionsfalse
\captionstyle{flushleft}

\begin{minipage}[b]{0.8\textwidth}
    \includegraphics[width=\textwidth,angle=-0]{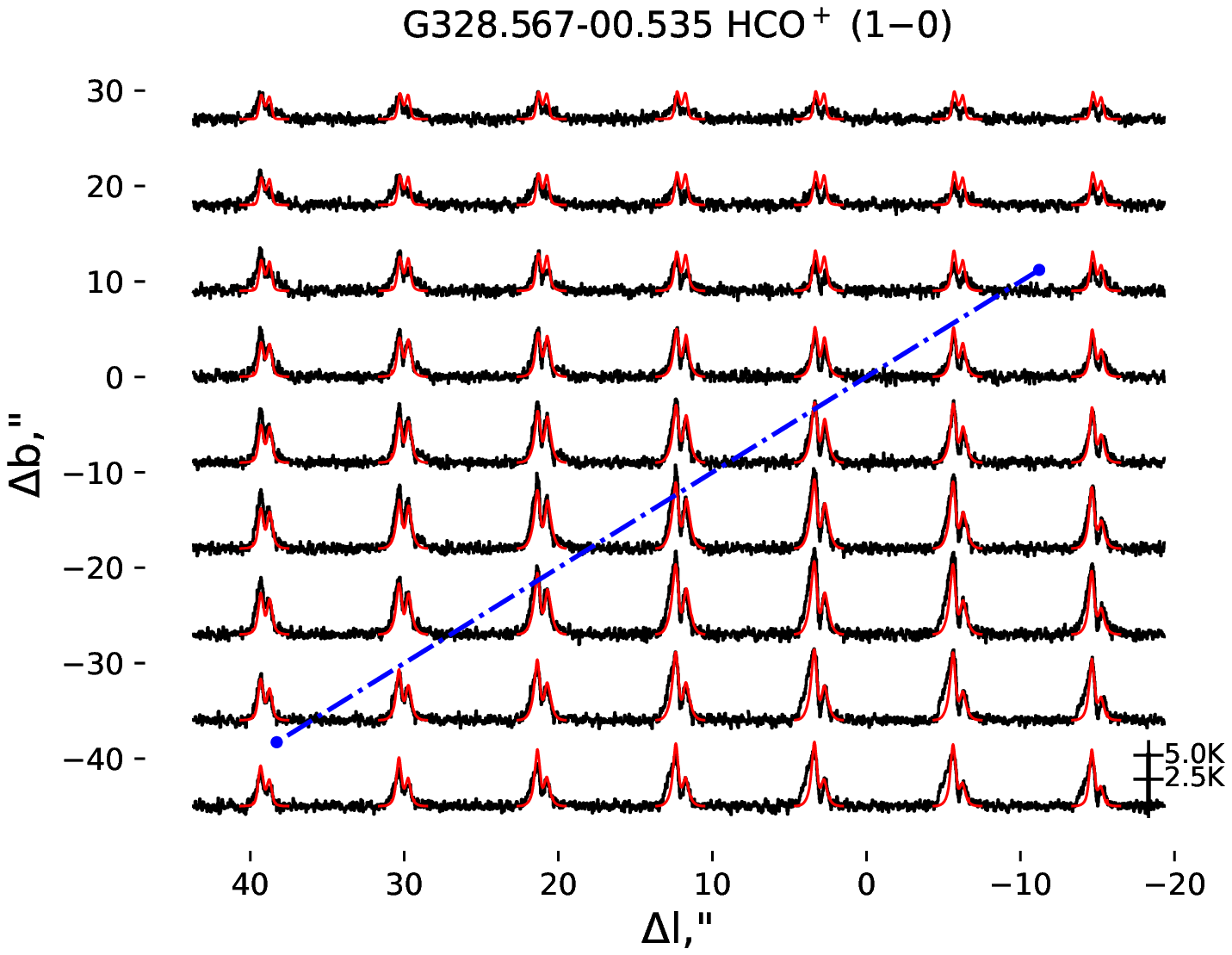}
\end{minipage}

\vskip 3mm

\begin{minipage}[b]{0.8\textwidth}
    \includegraphics[width=\textwidth,angle=-0]{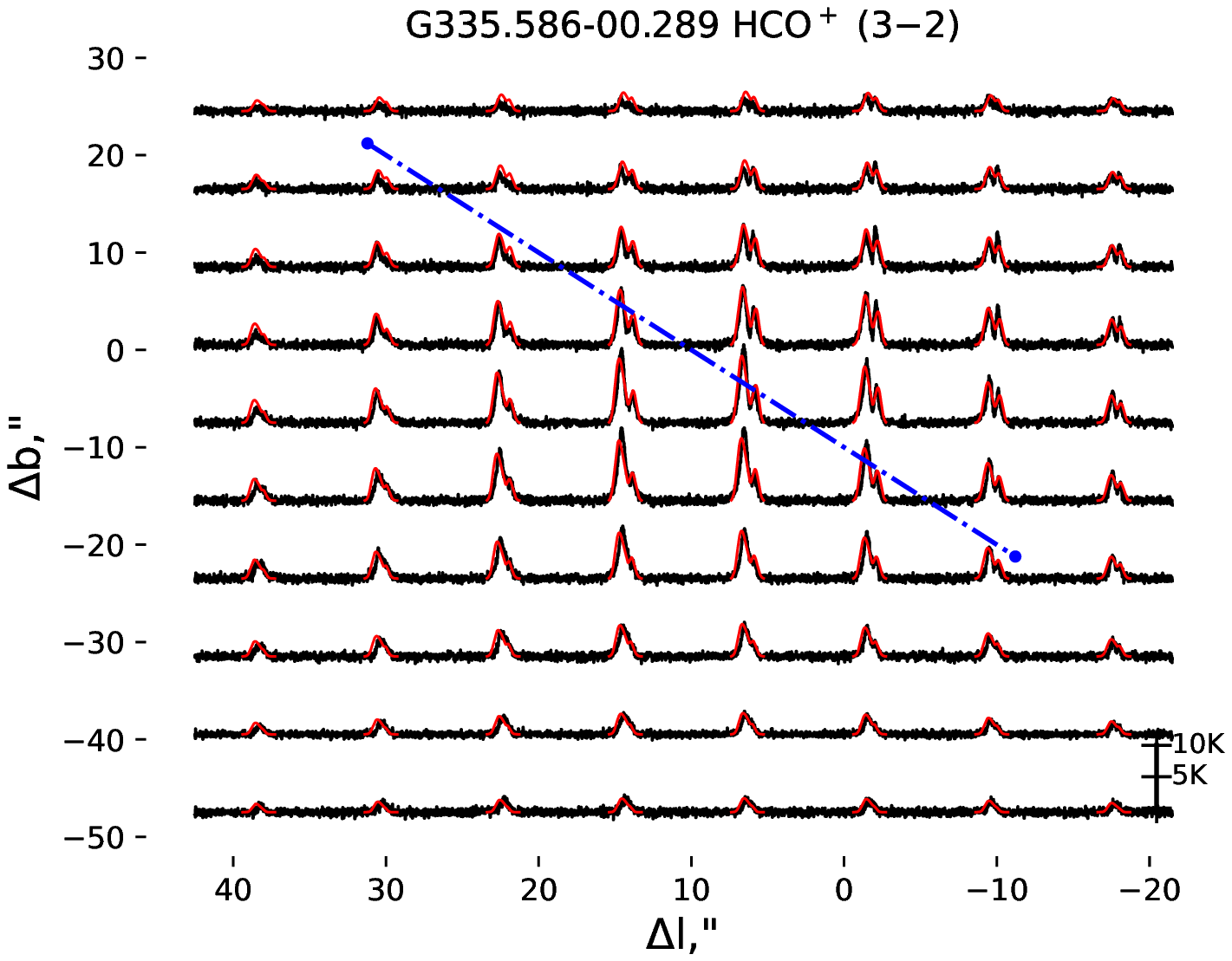}
\end{minipage}

\caption{\scriptsize
Observed (black) and model (red) (1--0) and (3--2) HCO$^+$ maps of G328.567 
and  G335.586, respectively. 
The velocity varies from --65 to --30~km/s and from --55 to -35~km/s 
for G328.567 and G335.586, respectively. 
Here, the model with radial and rotational motions is used.
The blue lines show the rotation axes. Above the axes,
the rotational motions are directed toward the observer, below them, 
away from the observer.
}
\label{fig:rot}
\end{figure}

\subsection{Velocity Profiles in the Gas Surrounding the Cores}

One of the results of our analysis was that the contraction
velocities of the cores are practically independent
of the radial distance. This fact may be related to
the fact that the cores are not isolated objects, but are
in a state of global contraction, and gas is probably
supplied to them from the surrounding gas. To search
for indications of the existence of such processes, it is
necessary to analyze the kinematics of the surrounding
gas based on observations in lines excited at lower densities
and on scales exceeding the scale of the cores.

Three cores from our sample (G012.418+00.506,
G335.586--00.289, and G343.127--00.063) were
observed in the SEDIGISM survey in the $^{13}$CO(2--1)
and C$^{18}$O(2--1) lines.
\footnote{The SEDIGISM database (https://sedigism.mpifrbonn.
mpg.de/index.html) was created by James Urquhart and is
maintained by the Max Planck Institute for Radio Astronomy.}
The $^{13}$CO(2--1) line is an indicator
of gas with a lower density than HCO$^+$(1--0)
($\sim 10^4$~cm$^{-3}$). 
Figure~5 shows maps of the regions containing
the cores in the $^{13}$CO(2--1) line (left panels),
the velocity profiles calculated from the $^{13}$CO(2--1)
lines, and the integrated intensity profiles (right panels).
The maps mark the positions for which the
$^{13}$CO(2--1) line velocities were calculated. The
$^{13}$CO(2--1) line velocities are in most cases close to
the corresponding velocities of the optically thin
C$^{18}$O(2--1) lines. It turned out that in all cases a
decrease in velocity (down to $\ga 1$~km/s) is observed in
the direction of the cores compared to the velocities of
the surrounding gas (the ``V''--type velocity profile).
This may be due to flows of the surrounding gas
toward the core located on the line of sight closer to
the observer (see, for example, [77, 78]). 
The right panels of Fig.~5 show with smooth curves the results 
of fitting into the observations data the projections of 
the radial profile of the gas velocity in the filament along
which the flow is directed: 
$V(L)=V_{LSR}-V_{sys}(L)\,cosA$ (see, for example, [77]). 
In the given formula, $L=r\,sinA$ is the projection of the
radial distance $r$ onto the plane of the sky, $A$ is the
angle between the direction of the radial gas motions
and the line of sight. The radial velocity profile was
specified as $V_{sys}\propto r^{-\alpha}$. The curves correspond to
three different values of $\alpha$. The model profiles for
$\alpha\la 0.5$ for all three cores describe the data better
compared to $\alpha=1$, while the model with $\alpha=0.1$ has
a slightly smaller residual and, thus, may be preferable
compared to $\alpha=0.5$. The analysis of the velocity profiles
constructed in the directions perpendicular to
those shown in Fig.~5 did not reveal any ``V''--type features
for G012.418 and G335.586, which indicates an
anisotropic nature of the gas movements from the outside
to these cores. In the core of G343.127, such a feature
is also noted along the perpendicular direction;
the gas flows here may be more isotropic.

Summarizing the obtained results, we note that for
all five massive cores associated with the regions of
formation of massive stars at different stages of evolution,
the power-law index of the radial profile of the
contraction velocity turned out to be practically flat,
differing from the value of 0.5 corresponding to the
regime of free fall of gas onto the protostar. The velocity
profiles in the surrounding gas of lower density, calculated
from the $^{13}$CO(2--1) data in the cores of
G012.418, G335.586 and G343.127, indicate the existence
of flows, the velocity profiles of which have indices $\la 0.5$. 
Apparently, the cores are nonequilibrium
and interact with the surrounding gas. This is consistent
with the conclusions of [27, 47] on the global collapse
of the cores of G012.418 and G335.586. For
these cores, flows from the outside apparently occur
along large-scale filaments, while for the core of
G343.127 the inflow of gas from the outside can be
carried out relatively isotropically. Thus, the obtained
results correspond to theoretical models considering
cores as nonequilibrium objects in a state of global collapse
(see, for example, [3, 7]). To confirm the
obtained conclusions about the kinematics of gas in
cores and the gas surrounding them, as well as to
obtain more rigorous estimates of physical parameters,
further studies are needed on various scales of
both already studied cores and new objects associated
with regions of formation of massive stars, with better
angular resolution and better sensitivity.

\begin{figure}[t!]
\setcaptionmargin{5mm}
\onelinecaptionsfalse
\captionstyle{flushleft}

\begin{minipage}[b]{0.4\textwidth}
    \includegraphics[width=\textwidth,angle=-0]{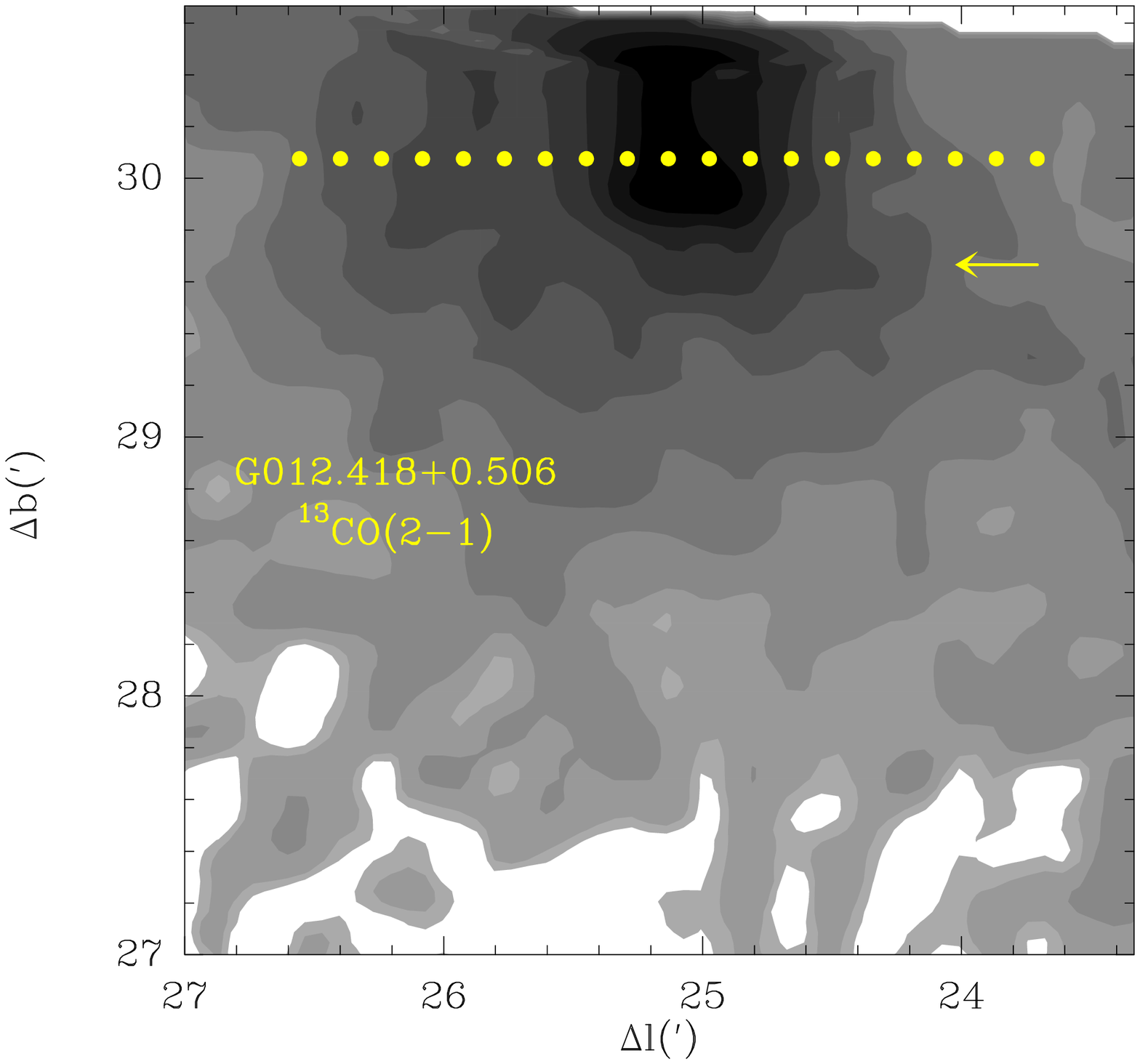}
\end{minipage}
\hspace{3mm}
\begin{minipage}[b]{0.365\textwidth}
    \includegraphics[width=\textwidth,angle=-0]{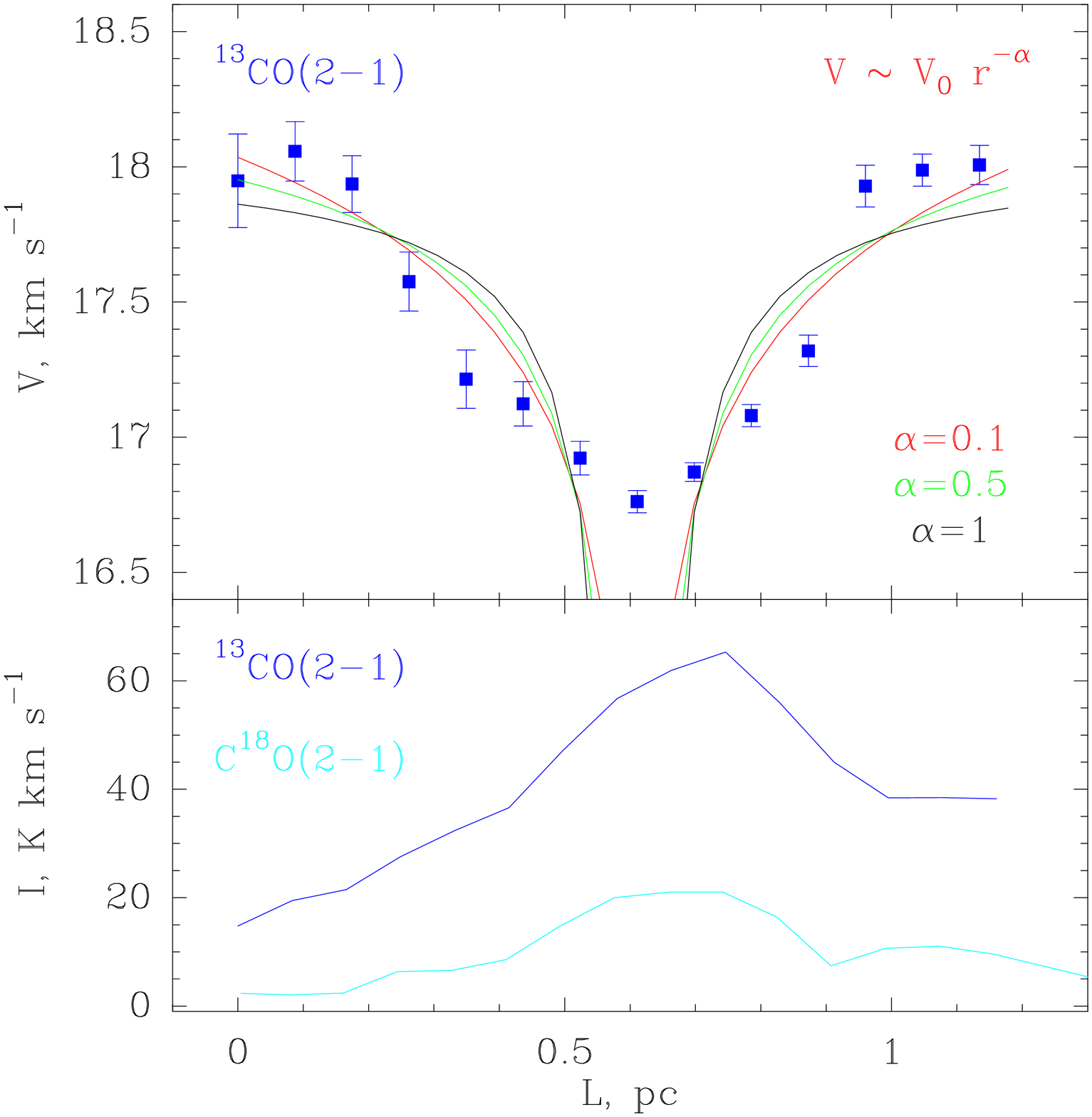}
\end{minipage}

\vspace{3mm}

\begin{minipage}[b]{0.4\textwidth}
    \includegraphics[width=\textwidth,angle=-0]{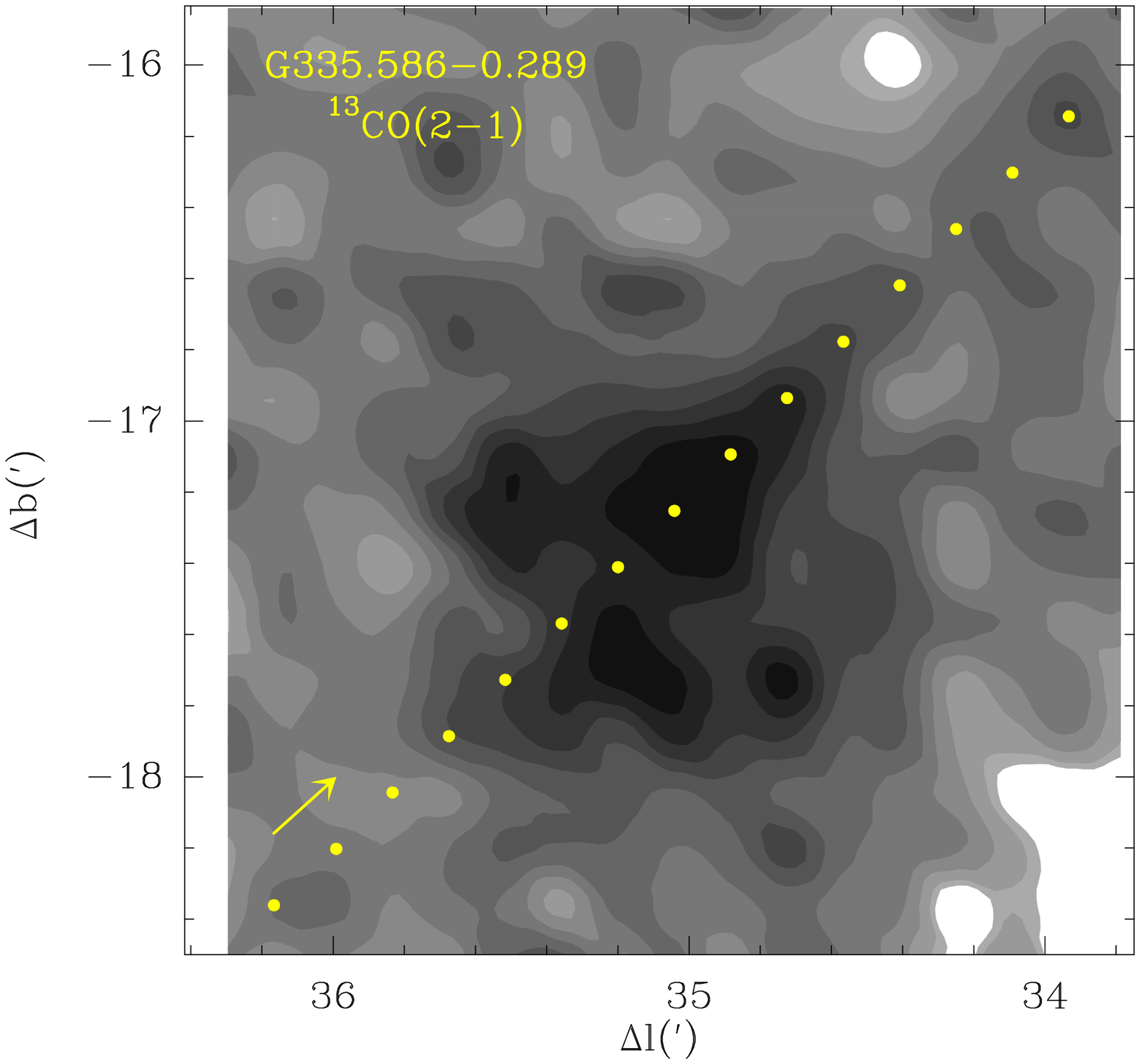}
\end{minipage}
\hspace{3mm}
\begin{minipage}[b]{0.365\textwidth}
    \includegraphics[width=\textwidth,angle=-0]{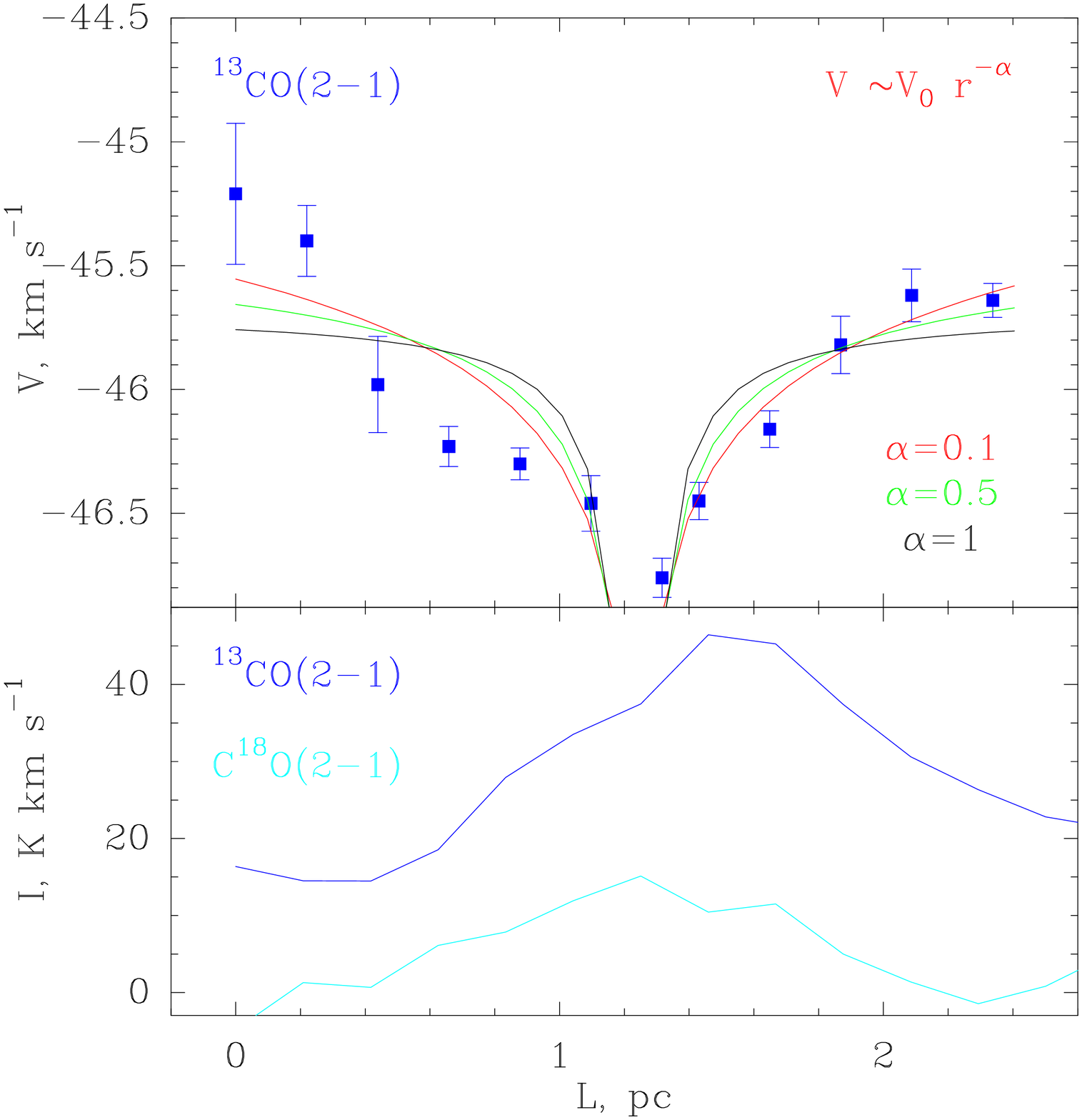}
\end{minipage}

\vspace{3mm}

\begin{minipage}[b]{0.4\textwidth}
    \includegraphics[width=\textwidth,angle=-0]{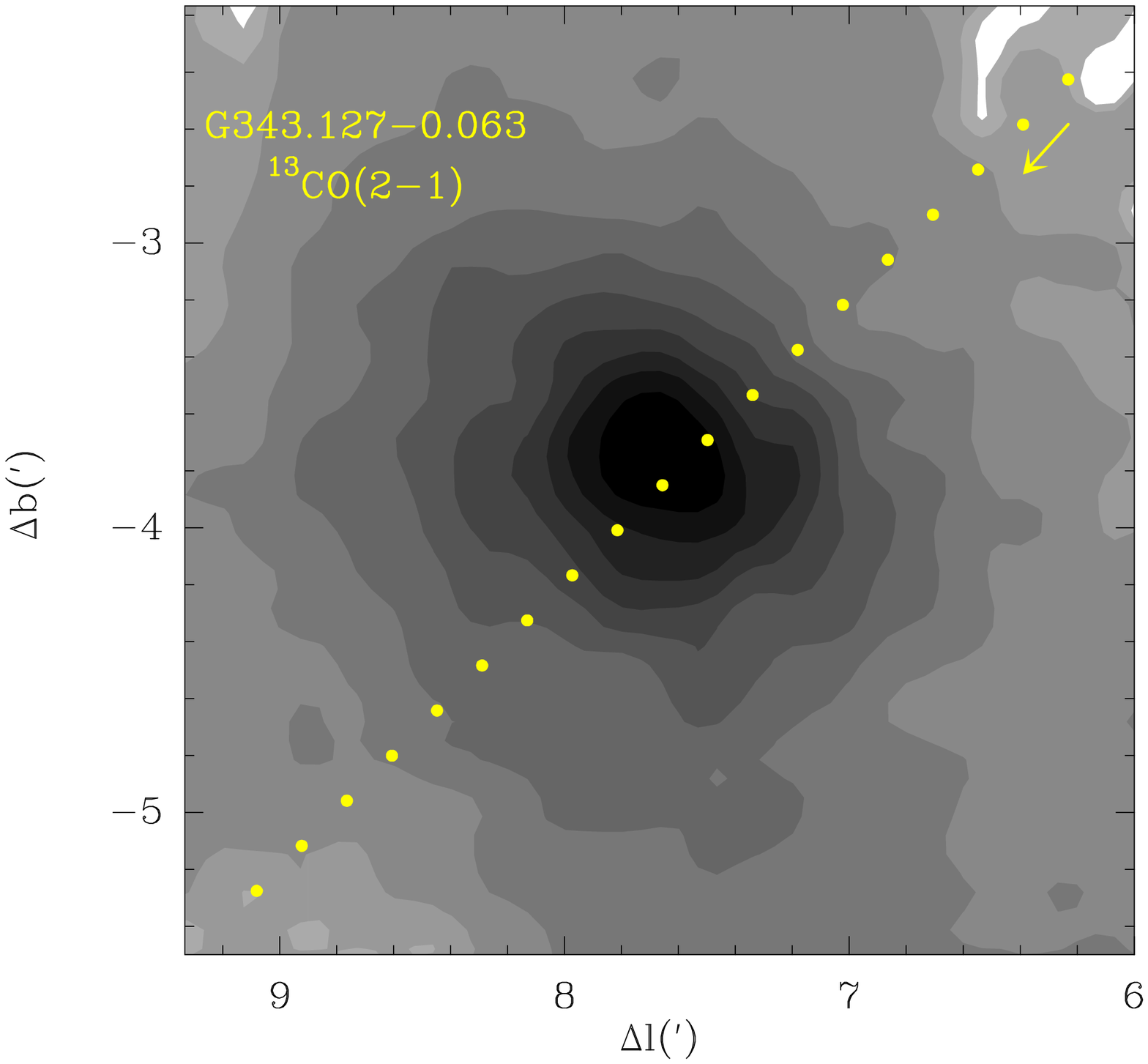}
\end{minipage}
\hspace{3mm}
\begin{minipage}[b]{0.365\textwidth}
    \includegraphics[width=\textwidth,angle=-0]{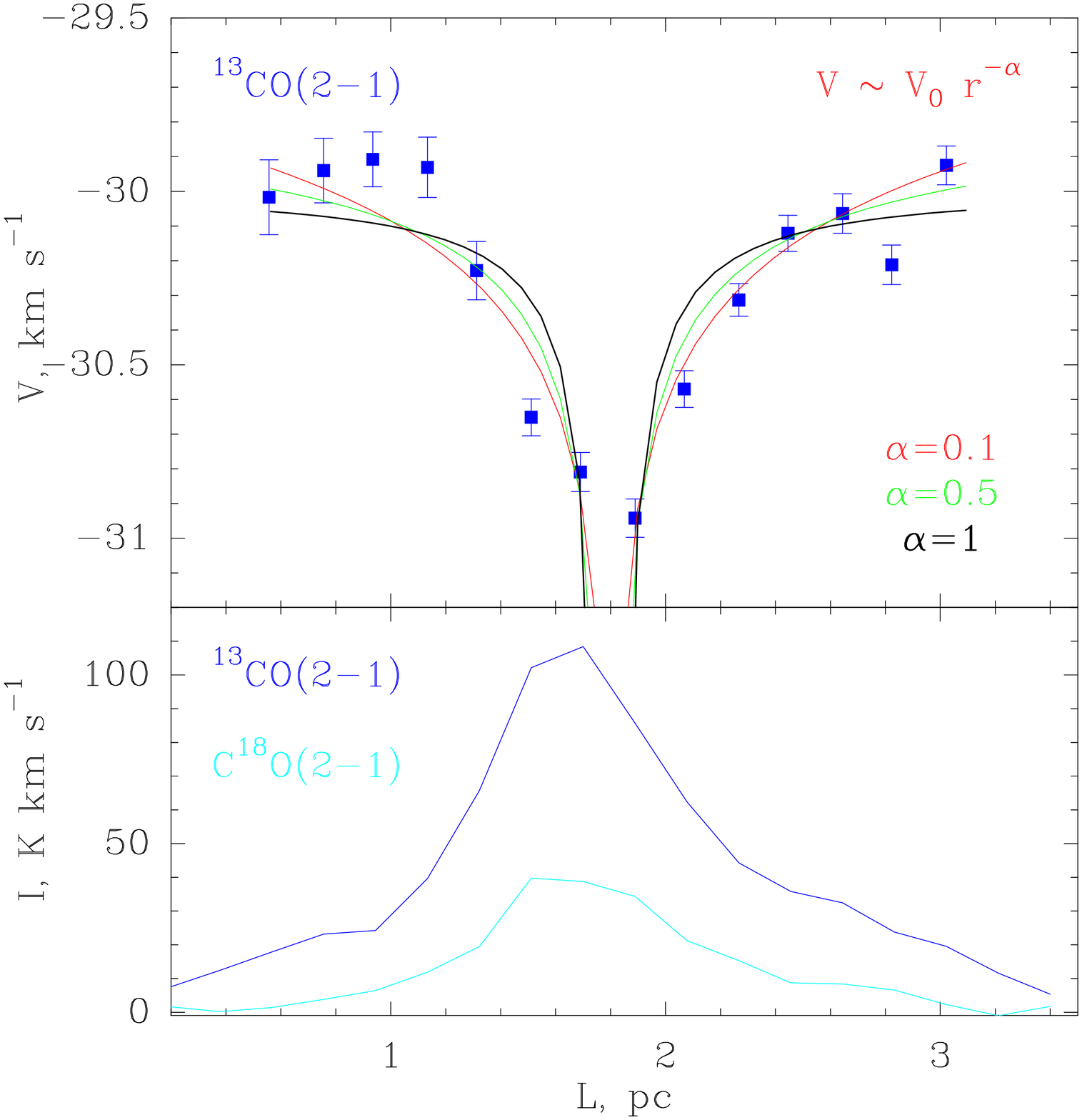}
\end{minipage}

\caption{\scriptsize
Maps of integrated $^{13}$CO(2--1) intensities for three cores from the sample 
based on SEDIGISM data [17] (left). 
The axes show the offsets relative to the galactic coordinates of the region 
centers: (12$^o$, 0$^o$), (335$^o$, 0$^o$), (343$^o$, 0$^o$), respectively. 
The maps show the positions for which the velocities of the line centers and 
their errors were calculated from the Gaussian approximation
of the $^{13}$CO(2--1) lines. The arrows indicate the directions along which 
the velocity profiles and integrated intensities of $^{13}$CO(2--1) and 
C$^{18}$O(2--1) were calculated. These profiles, converted from angular units 
to linear units taking into account the distances
to the objects (Table~1), are shown in the graphs on the right. 
Smooth curves of different colors correspond to the results of fitting
the model dependences of the velocity on the distance 
with different power exponents.
}
\label{fig:vprof}
\end{figure}

\section{Conclusions}

In order to study the characteristics of contracting
massive cores, we analyzed the spectral observation
data of the objects G012.418+00.506,
G326.472+00.888, G328.567--00.535, G335.586--00.289, 
and G343.127--00.063 from the MALT90 database. 
The objects have dense cores associated with
regions of formation of massive stars at different stages
of evolution. The core of G326.472+00.888, for which
there are no indications of the existence of internal
sources, is apparently at the earliest stage of evolution
compared to the others. Optically thick HCO$^+$(1--0)
lines in all cores exhibit a ``blue'' asymmetry and a dip,
while lines with smaller optical depth (H$^{13}$CO$^+$(1--0),
N$_2$H$^+$(1--0)) are relatively symmetrical, which may
indicate contraction.

Data analysis by fitting spectral maps calculated
within the spherically symmetric model of the core
into the observed maps using the error function minimization
algorithm [9] allowed us to calculate the
optimal values of the parameters of the radial density,
turbulent velocity, and contraction velocity profiles in
the cores. The power-law index of density decay with
distance from the center varies in the range ,
the lowest value of the index was obtained for the core
G326.472+00.888. The turbulent velocity decreases
with the index $\sim 0.3-0.6$. The contraction velocity of
the cores depends weakly on the distance from the
center, the power-law index of contraction velocity
decay with distance from the center was $\sim 0.1$ for all
cores, which differs from the value of 0.5 for the free
fall mode. In the cores of G328.567--00.535 and
G335.586--00.289, in addition to radial motions, there
are indications of rotation. For G328.567--00.535, the
rotation law is close to Keplerian, for 
G335.586--00.289 the rotation velocity profile is close to flat.

Using the $^{13}$CO(2--1) data from the SEDIGISM
survey, velocity profiles were plotted for the three
regions. In the direction of the cores, ``V''--type features
were found that may be associated with gas flows
onto the core. Fitting the projections of the function of
the form $r^{-\alpha}$ into the obtained diagrams indicates the
preference of profiles with indices $\alpha\la 0.5$.

The results obtained indicate that the massive cores
under consideration interact with their environment
and are apparently in a state of global collapse.

\newpage

{\bf REFERENCES}

\vskip 2mm

1. J. C. Tan, M. T. Beltran, P. Caselli, F. Fontani,
A. Fuente, M. R. Krumholz, C. F. McKee, and A. Stolte,
in Protostars and Planets VI, Ed. by H. Beuther,
R. S. Klessen, C. P. Dullemond, and T. Henning
(Univ. Arizona Press, Tucson, 2014), p. 149.

2. F. Motte, S. Bontemps, and F. Louvet, Ann. Rev. Astron.
Astrophys. 56, 41 (2018).

3. A. Whitworth and D. Summers, Mon. Not. R. Astron.
Soc. 214, 1 (1985).

4. F. H. Shu, Astrophys. J. 214, 488 (1977).

5. C. F. McKee and J. C. Tan, Astrophys. J. 585, 850
(2003).

6. I. A. Bonnell, M. R. Bate, C. J. Clarke, and J. E. Pringle,
Mon. Not. R. Astron. Soc. 323, 785 (2001).

7. E. Vazquez-Semadeni, A. Palau, J. Ballesteros-Paredes,
G. C. Gomez, and M. Zamora-Aviles, Mon. Not.
R. Astron. Soc. 490, 3061 (2019).

8. P. C. Myers, D. Mardones, M. Tafalla, J. P. Williams,
and D. J. Wilner, Astrophys. J. Lett. 465, L133 (1996).

9. L. E. Pirogov and P. M. Zemlyanukha, Astron. Rep.
65, 82 (2021).

10. Y.-X. He, J.-J. Zhou, J. Esimbek, W.-G. Ji, et al., Mon.
Not. R. Astron. Soc. 461, 2288 (2016).

11. L. E. Pirogov, V. M. Shul'ga, I. I. Zinchenko, P. M. Zemlyanukha,
A. H. Patoka, and M. Tomasson, Astron.
Rep. 60, 904 (2016).

12. H. Yoo, K.-T. Kim, J. Cho, M. Choi, J. Wu, N. J. Evans II,
and L. M. Ziurys, Astrophys. J. Suppl. 235, 31 (2018).

13. L. E. Pirogov, P. M. Zemlyanukha, E. M. Dombek,
and M. A. Voronkov, Astron. Rep. 67, 1355 (2023).

14. J. M. Jackson, J. M. Rathborne, J. B. Foster, J. S. Whitaker,
et al., Publ. Astron. Soc. Austral. 30, e057 (2013).

15. F. Schuller, K. M. Menten, Y. Contreras, F. Wyrowski,
et al., Astron. Astrophys. 504, 415 (2009).

16. S. Molinari, B. Swinyard, J. Bally, M. Barlow, et al.,
Publ. Astron. Soc. Pacif. 122 (889), 314 (2010).

17. F. Schuller, J. S. Urquhart, T. Csengeri, D. Colombo,
et al., Mon. Not. R. Astron. Soc. 500, 3064 (2021).

18. T. Umemoto, T. Minamidani, N, Kuno, S. Fujita,
et al., Publ. Astron. Soc. Jpn. 69, 78 (2017).

19. J. M. Rathborne, J. S. Whitaker, J. M. Jackson,
J. B. Foster, et al., Publ. Astron. Soc. Austral. 33, e030
(2016).

20. N. J. Evans II, Ann. Rev. Astron. Astrophys. 37, 311
(1999).

21. J. S. Urquhart, M. R. A. Wells, T. Pillai, S. Leurini,
et al., Mon. Not. R. Astron. Soc. 510, 3389 (2022).

22. A. E. Guzman, P. Sanhueza, Y. Contreras, H. A. Smith,
J. M. Jackson, S. Hoq, and J. M. Rathborne, Astrophys.
J. 815, 130 (2015).

23. Galactic Legacy Infrared Midplane Survey Extraordinaire
(GLIMPSE). https://irsa.ipac.caltech.edu/data/
SPITZER/GLIMPSE/.

24. N. Issac, A. Tej, T. Liu, W. Varricatt, S. Vig, C. H. Ishwara
Chandra, and M. Schultheis, Mon. Not. R. Astron.
Soc. 485, 1775 (2019).

25. C. J. Cyganowski, B. A. Whitney, E. Holden, E. Braden,
et al., Astron. J. 136, 2391 (2008).

26. X. Chen, Z.-Q. Shen, J.-J. Li, Y. Xu, and J.-H. He, Astrophys.
J. 710, 150 (2010).

27. A. Saha, A. Tej, H.-L. Liu, T. Liu, et al., Mon. Not. R.
Astron. Soc. 516, 1983 (2022).

28. J. S. Urquhart, M. G. Hoare, C. R. Purcell,
S. L. Lumsden, et al., Astron. Astrophys. 501, 539
(2009).

29. Y. L. Shirley, N. J. Evans II, K. E. Young, C. Knez, and
D. T. Jaffe, Astrophys. J. Suppl. 149, 375 (2003).

30. C. J. Cyganowski, J. Koda, E. Rosolowsky, S. Towers,
J. Donovan Meyer, F. Egusa, R. Momose, and
T. P. Robitaille, Astrophys. J. 764, 61 (2013).

31. R. Cesaroni, F. Palagi, M. Felli, M. Catarzi, G. Comoretto,
S. di Franco, C. Giovanardi, and F. Palla,
Astron. Astrophys. Suppl. Ser. 76, 445 (1988).
32. B. E. Svoboda, Y. L. Shirley, C. Battersby, E. W. Rosolowsky,
et al., Astrophys. J. 822, 59 (2016).

33. X. Chen, S. P. Ellingsen, Z.-Q. Shen, A. Titmarsh, and
C.-G. Gan, Astrophys. J. Suppl. 196, 9 (2011).

34. W. Yang, Y. Xu, X. Chen, S. P. Ellingsen, D. Lu, B. Ju,
and Y. Li, Astrophys. J. Suppl. 231, 20 (2017).

35. H. Nguyen, M. R. Rugel, C. Murugeshan, K. M. Menten,
et al., Astron. Astrophys. 666, A59 (2022).

36. D. A. Ladeyschikov, O. S. Bayandina, and A. M. Sobolev,
Astron. J. 158, 233 (2019).

37. T. Csengeri, S. Bontemps, F. Wyrowski, S. T. Megeath,
F. Motte, A. Sanna, M. Wienen, and K. M. Menten,
Astron. Astrophys. 601, A60 (2017).

38. T. Mauch, T. Murphy, H. J. Buttery, J. Curran,
R. W. Hunstead, B. Piestrzynski, J. G. Robertson, and
E. M. Sadler, Mon. Not. R. Astron. Soc. 342, 1117
(2003).

39. N.-P. Yu, J.-L. Xu, J.-J. Wang, and X.-L. Liu, Astrophys.
J. 865, 135 (2018).

40. Y. Lin, T. Csengeri, F. Wyrowski, J. S. Urquhart,
F. Schuller, A. Weiss, and K. M. Menten, Astron. Astrophys.
631, A72 (2019).

41. T. Culverhouse, P. Ade, J. Bock, M. Bowden, et al.,
Astrophys. J. Suppl. 195, 8 (2011).

42. J. P. Phillips and J. A. Perez-Grana, Mon. Not. R. Astron.
Soc. 393, 354 (2009).

43. F. Fontani, M. T. Beltran, J. Brand, R. Cesaroni,
L. Testi, S. Molinari, and C. M. Walmsley, Astron. Astrophys.
432, 921 (2005).

44. G. C. MacLeod, D. J. van der Walt, A. North,
M. J. Gaylard, J. A. Galt, and G. H. Moriarty-
Schieven, Astron. J. 116, 2936 (1998).

45. R. J. Cohen, M. R. W. Masheder, and J. L. Caswell,
Mon. Not. R. Astron. Soc. 274, 808 (1995).

46. N. Peretto and G. A. Fuller, Astron. Astrophys. 505,
405 (2009).

47. F.-W. Xu, K. Wang, T. Liu, P. F. Goldsmith, et al.,
Mon. Not. R. Astron. Soc. 520, 3259 (2023).

48. K. Ishihara, P. Sanhueza, F. Nakamura, M. Saito,
et al., arXiv: 2407.06845 [astro-ph.GA] (2024).

49. M. Anderson, N. Peretto, S. E. Ragan, A. J. Rigby,
et al., Mon. Not. R. Astron. Soc. 508, 2964 (2021).

50. A. J. Walsh, C. R. Purcell, S. N. Longmore, S. L. Breen,
J. A. Green, L. Harvey-Smith, C. H. Jordan, and
C. Macpherson, Mon. Not. R. Astron. Soc. 442, 2240
(2014).

51. S. L. Breen, J. L. Caswell, S. P. Ellingsen, and
C. J. Phillips, Mon. Not. R. Astron. Soc. 406, 1487
(2010).

52. J. L. Caswell, J. A. Green, and C. J. Phillips, Mon. Not.
R. Astron. Soc. 439, 1680 (2014).

53. M. A. Voronkov, J. L. Caswell, S. P. Ellingsen,
J. A.Green, and S. L. Breen, Mon. Not. R. Astron.
Soc. 439, 2584 (2014).

54. S. L. Breen, Y. Contreras, J. R. Dawson, S. P. Ellingsen,
M. A. Voronkov, and T. P. McCarthy, Mon.
Not. R. Astron. Soc. 484, 5072 (2019).

55. X. Chen, S. P. Ellingsen, Z.-Q. Shen, A. Titmarsh, and
C.-G. Gan, Astrophys. J. Suppl. 196, 9 (2011).

56. A. J. Walsh, M. G. Burton, A. R. Hyland, and G. Robinson,
Mon. Not. R. Astron. Soc. 301, 640 (1998).

57. J. L. Caswell, Publ. Astron. Soc. Austral. 26, 454
(2009).

58. G. Garay, D. Mardones, L. Bronfman, K. J. Brooks,
et al., Astron. Astrophys. 463, 217 (2007).

59. G. Garay, K. J. Brooks, D. Mardones, and R. P. Norris,
Astrophys. J. 587, 739 (2003).

60. L. F. Rodriguez, G. Garay, K. J. Brooks, and D. Mardones,
Astrophys. J. 626, 953 (2005).

61. L. A. Zapata, G. Garay, A. Palau, L. F. Rodriguez,
M. Fernandez-Lopez, R. Estalella, and A. Guzman,
Astrophys. J. 872, 176 (2019).

62. J. R. Forster and J. L. Caswell, Astron. Astrophys. Suppl.
Ser. 137, 43 (1999).

63. H.-H. Qiao, A. J. Walsh, J. A. Green, S. L. Breen,
et al., Astrophys. J. Suppl. 227 (2), 26 (2016).

64. V. I. Slysh, S. V. Kalenskii, I. E. Valtts, and R. Otrupcek,
Mon. Not. R. Astron. Soc. 268, 464 (1994).

65. L. Pirogov, I. Zinchenko, P. Caselli, and L. E. B. Johansson,
Astron. Astrophys. 461, 523 (2007).

66. L. Pirogov, I. Zinchenko, P. Caselli, L. E. B. Johansson,
and P. C. Myers, Astron. Astrophys. 405, 639
(2003).

67. Y. Pavlyuchenkov, D. Wiebe, R. Launhardt, and T. Henning,
Astrophys. J. 645, 1212 (2006).

68. S. Terebey, C. J. Chandler, and P. Andre, Astrophys. J.
414, 759 (1993).

69. S. D. Doty and C. M. Leung, Astrophys. J. 424, 729
(1994).

70. Y. N. Pavlyuchenkov, D. S. Wiebe, A. M. Fateeva, and
T. S. Vasyunina, Astron. Rep. 55, 1 (2011).

71. Y. N. Pavlyuchenkov, A. G. Zhilkin, E. I. Vorobyov,
and A. M. Fateeva, Astron. Rep. 59, 133 (2015).

72. D. R. Flower, Mon. Not. R. Astron. Soc. 305, 651
(1999).

73. M. Juvela, Astron. Astrophys. 644, A151 (2020).

74. L. E. Pirogov, Astron. Rep. 53, 1127 (2009).

75. S. A. Khaibrakhmanov, A. E. Dudorov, N. S. Kargaltseva,
and A. G. Zhilkin, Astron. Rep. 65, 693 (2021).

76. M. R. A. Wells, H. Beuther, S. Molinari, P. Schilke,
et al., Astron. Astrophys. 690, A185 (2024).

77. A. Hacar, J. Alves, M. Tafalla, and J. R. Goicoechea,
Astron. Astrophys. 602, L2 (2017).

78. J. W. Zhou, F. Wyrowski, S. Neupane, J. S. Urquhart,
et al., Astron. Astrophys. 676, A69 (2023).

\vskip 3mm

\hspace{10cm} \it{Translated by T.\,Sokolova}

\end{document}

%% file: AADEF.TEX
\def\la{\mathrel{\mathchoice {\vcenter{\offinterlineskip\halign{\hfil
$\displaystyle##$\hfil\cr<\cr\sim\cr}}}
{\vcenter{\offinterlineskip\halign{\hfil$\textstyle##$\hfil\cr
<\cr\sim\cr}}}
{\vcenter{\offinterlineskip\halign{\hfil$\scriptstyle##$\hfil\cr
<\cr\sim\cr}}}
{\vcenter{\offinterlineskip\halign{\hfil$\scriptscriptstyle##$\hfil\cr
<\cr\sim\cr}}}}}
\def\ga{\mathrel{\mathchoice {\vcenter{\offinterlineskip\halign{\hfil
$\displaystyle##$\hfil\cr>\cr\sim\cr}}}
{\vcenter{\offinterlineskip\halign{\hfil$\textstyle##$\hfil\cr
>\cr\sim\cr}}}
{\vcenter{\offinterlineskip\halign{\hfil$\scriptstyle##$\hfil\cr
>\cr\sim\cr}}}
{\vcenter{\offinterlineskip\halign{\hfil$\scriptscriptstyle##$\hfil\cr
>\cr\sim\cr}}}}}

\def\utw{\smash{\rlap{\lower5pt\hbox{$\sim$}}}}
\def\udtw{\smash{\rlap{\lower6pt\hbox{$\approx$}}}}

\def\diameter{{\ifmmode\mathchoice
{\ooalign{\hfil\hbox{$\displaystyle/$}\hfil\crcr
{\hbox{$\displaystyle\mathchar"20D$}}}}
{\ooalign{\hfil\hbox{$\textstyle/$}\hfil\crcr
{\hbox{$\textstyle\mathchar"20D$}}}}
{\ooalign{\hfil\hbox{$\scriptstyle/$}\hfil\crcr
{\hbox{$\scriptstyle\mathchar"20D$}}}}
{\ooalign{\hfil\hbox{$\scriptscriptstyle/$}\hfil\crcr
{\hbox{$\scriptscriptstyle\mathchar"20D$}}}}
\else{\ooalign{\hfil/\hfil\crcr\mathhexbox20D}}%
\fi}}